\documentclass[a4paper,12pt]{article}

\usepackage[utf8]{inputenc}
\usepackage{graphicx}        
\usepackage{amsmath,mathrsfs}

\usepackage{mathtools}
\usepackage{amssymb} 

\usepackage{cancel}

\usepackage{empheq}

\newcounter{mnotecount}

\newcommand{\mnotex}[1]
{\protect{\stepcounter{mnotecount}}$^{\mbox{\footnotesize $\bullet$\themnotecount}}$ 
\marginpar{
\raggedright\tiny\em
$\!\!\!\!\!\!\,\bullet$\themnotecount: #1} }

\newcommand{\bm}[1]{\mbox{\boldmath $#1$}}
\def\defi{:=}

\def \a {\alpha}
\def \b {\beta}
\def \m {\mu}
\def \n {\nu}
\def \r {\rho}
\def \g {\gamma}
\def \l {\lambda}

\def \d {\delta}

\def \N {\nabla}
\def \Nb {\overline{\nabla}}
\def \Boxb {\overline\Box}

\def \Rb {\overline{R}}

\def \au {a_1}
\def \ad {a_2}
\def \at {a_3}

\def\deltaS{\delta^\Sigma}

\def \dl {\underline\Delta}

\def\RS{R^\Sigma}

\def\S{\Sigma}

\def\otheta{\underline{\theta}}
\def\1{\underline{1}}

\def\be{\begin{equation}}
\def\ee{\end{equation}}
\def\bea{\begin{eqnarray}}
\def\eea{\end{eqnarray}}
\def\bean{\begin{eqnarray*}}
\def\eean{\end{eqnarray*}}

\newtheorem{result}{Result}[section]

\newtheorem{definition}{Definition}[section]

\begin{document}
\title{Junction conditions in quadratic gravity:\\ thin shells and double layers
}
\author{Borja Reina, Jos\'e M. M. Senovilla, and Ra\"ul Vera \\
F\'{\i}sica Te\'orica, Universidad del Pa\'{\i}s Vasco, \\
Apartado 644, 48080 Bilbao, Spain \\ 
e-mail: josemm.senovilla@ehu.es \\
raul.vera@ehu.es \\
borja.reina@ehu.es} 
\date{}
\maketitle
\begin{abstract} 
The junction conditions for the most general gravitational theory with
a Lagrangian containing terms quadratic in the curvature are
derived. We include the cases with a possible concentration of matter
on the joining hypersurface ---termed as thin shells, domain walls or
braneworlds in the literature--- as well as the proper matching
conditions where only finite jumps of the energy-momentum tensor are
allowed. In the latter case we prove that the matching conditions are
more demanding than in General Relativity. In the former case, we show
that generically the shells/domain walls are of a new kind because
they possess, in addition to the standard energy-momentum tensor, a
{\em double layer} energy-momentum contribution which actually induces
an external energy flux vector and an external scalar pressure/tension
on the shell.  We prove that all these contributions are necessary to
make the entire energy-momentum tensor divergence-free, and we present
the field equations satisfied by these energy-momentum quantities. The
consequences of all these results are briefly analyzed.
\end{abstract} 


\section{Introduction}
{\em Quadratic gravity} refers to theories generalizing General Relativity (GR) by adding terms quadratic in the curvature to the Lagrangian density. The motivations for such modifications go back several decades ago (see the critic paper \cite{Rose1991}), and today there is a general consensus that
modern string theory (see e.g. \cite{Gaume2015}) and other approaches to quantum gravity  (see e.g. \cite{Olmo2011})
present that structure, even with higher powers of the curvature tensor, in their effective actions.

On the other hand, many times it is convenient to have a description of concentrated
sources, that is, of concentrated matter and energy in gravity
theories. These concentrated sources represent for instance thin
shells of matter (or braneworlds, or domain walls) and impulsive
matter or gravitational waves. They can mathematically be modelled by
using distributions, such as Dirac deltas or the like, hence, one has
to resort to using {\em tensor distributions}.  However, one cannot
simply assume that the metric is a distribution because the products
of distributions is not well defined in general, and therefore the
curvature (and Einstein) tensor will not be defined. Thus, one must
identify the class of metrics whose curvature is defined as a
distribution, and such that the field equations make sense. For
sources on thin shells, the appropriate class of metrics were
identified in \cite{I,L,T} in GR, further discussed in
\cite{GT}. Essentially, these are the metrics which are smooth except
on localized hypersurfaces where the metric is only continuous.

We carry on a similar program in the most general quadratic theory of
gravity, where extra care must be taken: the field equations, as well
as the Lagrangian density, contain products of Riemann tensors, and,
moreover, their second derivatives.  Therefore, the {\em singular
distributional part} ---such as the Dirac deltas--- cannot arise in
the Riemann tensor itself, which can have at most finite jumps except in some very excepctional situations. We identify these and then concentrate on the generic, and more relevant, situation 
performing a detailed calculation using the rigorous calculus of tensor
distributions (see the Appendices for definitions and fundamental
formulas with derivations) to obtain the energy-momentum quantities on
the shells. They depend on the extrinsic geometrical properties of the
hypersurface supporting it, as well as on the possible discontinuities
of the curvature and their derivatives.

Surprisingly, and as already demonstrated in \cite{Senovilla13,Senovilla14,Senovilla15}, a
contribution of ``dipole'' type also appears in the energy-momentum
content supported on the shell. This is what we call a double layer,
in analogy with the terminology used in classical electrodynamics
\cite{J} for the case of electrodipole surface
distributions. This analogy make the interpretation of these double layers somewhat misterious, as there are no negative masses ---ad thus no mass dipoles--- in gravitation.
One of our purposes is to shed some light into this new mystery. From our results and those in \cite{Senovilla13,Senovilla14,Senovilla15}, these double layers seem to arise when abrupt changes in the Einstein tensor occur.

We also find the field equations obeyed by all these energy-momentum
quantities, which generalize the traditional Israel equations
\cite{I}, and describe the conservation of energy and
momentum. Actually, we explicitly prove that the full energy-momentum tensor
is divergence-free (in the distributional sense) by virtue of the
mentioned field equations.

Previous works on junction conditions in quadratic gravity include
\cite{BD,D,DSS,BF} ---see also \cite{DD,GW} for the Gauss-Bonnet
case---, but none of them provided the correct full field equations
with matter outside the shell, and they all missed the double-layer
contributions, which are fundamental for the energy-momentum
conservation. 
Maybe this is due to the
extended use of Gaussian coordinates based on the thin shell: this
prevents from making a mathematically sound analysis of the
distributional part of the energy-momentum tensor, as the derivatives
of the Dirac delta supported on the shell seem to be ill-defined in
those coordinates. This is explained in detail in Appendix
\ref{App:E}.

The paper
is structured as follows. In Section \ref{sec:matching} we present
a purely geometric review on spacetimes with distributional curvature constructed by joining
smooth spacetimes. The quadratic gravity field equations are introduced in
Section \ref{sec:quadratic_grav}, where the proper junction conditions for the
description of thin shells (layers) are found. This is achieved by using distributional calculus,
briefly reviewed in the Appendices. In Section \ref{sec:compute_deltas},
the matter content supported on the layer, i.e.
the distributional part of the global energy momentum tensor, is  found
to contain  a ``usual'' Dirac-delta term $\widetilde{T}_{\m\n}\deltaS$ together with another contribution of double-layer type as
mentioned above; the latter is denoted by $\underline{t}_{\mu\nu}$.
Then, both $\widetilde{T}_{\m\n}$ and $\underline{t}_{\mu\nu}$ are computed
in terms of geometrical quantities: the curvatures at either side of the layer and the extrinsic and intrinsic geometry of the hypersurface supporting it.
The tensor $\widetilde{T}_{\m\n}$ is decomposed into the proper energy momentum of the shell
$\tau_{\a\b}$, external flux momentum $\tau_\a$ and external pressure (or tension) $\tau$
corresponding to the  completely tangent, tangent-normal and normal parts respectively.
The double layer energy-momentum tensor distribution is found to resemble the energy-momentum content of
a dipole surface charge distribution with strength $\mu_{\a\b}$.
This strength depends on the jump of the Einstein --or equivalently the Ricci--- tensor at the layer.
The allowed jumps of the curvature (and its derivatives up to second order)
at the layer are determined in Section \ref{sec:nG2}, again from a purely geometrical perspective. 

The general quadratic gravity field equations are obtained in Section \ref{sec:field_eqs_S}. These are the inherited field equations on the layer, and they involve $\tau_{\a\b}$, $\tau_\a$, $\tau$
and $\mu_{\a\b}$ together with jumps on the layer of the spacetime energy-momentum tensor.  These fundamental equations are the generalization of the Israel equations in GR
to the general quadratic gravity theories. The covariant conservation of the full energy-momentum tensor with its distributional parts 
is explicitly demonstrated in Section \ref{section:divergence}, where we discuss how the double layer
term is necessary for that.
The field equations on the layer are analysed and further discussed in Section \ref{sec:8},
where a classification of the junction conditions in the following cases are presented:
proper matching, thin shells with no double layers, and pure double layers.
In particular we find that if there is no double layer, then no external flux momentum $\tau_\a$
nor external tension  $\tau$ can exist.
Finally, in Section \ref{sec:consequences} some comparisons with the general GR case,
and particular matchings of spacetimes, are provided.
It is found that any GR solution containing a proper matching
hypersurface will contain a double layer and/or a thin shell at the matching hypersurface
if the true theory is quadratic. Therefore, if any quantum
regimes require, excite or switch on quadratic terms in the Lagrangian density, then GR
solutions modelling two regions with different matter contents will develop thin shells and
double layers on their interfaces.

In order to have a self-contained text, we devote some Appendices to review
distributional calculus in manifolds and to present some
useful general calculations with distributions. On the other hand,
we present in Appendix \ref{App:E} a (we hope clarifying)
discussion about the difficulty, and in fact inconvenience,
of using Gaussian coordinates for dealing with layers in quadratic Lagrangian
theories, as it has been often done in the literature.

\section{Junction: spacetimes with distributional curvature}\label{sec:matching}
\begin{figure}[!ht]
\includegraphics[height=8cm]{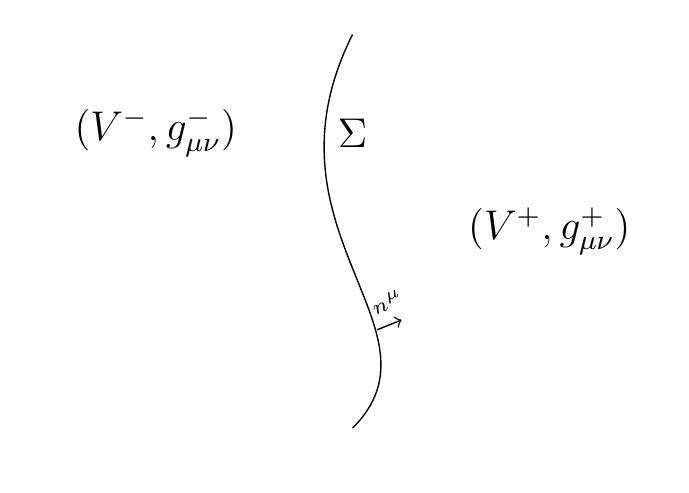}
\caption{\footnotesize{Schematic diagram of the situation under consideration: $\S$ is a timelike hypersurface separating two regions of the space-time, $V^+$ and $V^-$, with corresponding smooth metrics $g^+$ and $g^-$. These two metrics also have well defined, definite limits, when approaching $\S$. If, and only if, the first fundamentals forms inherited by $\S$ from $V^+$ and $V^-$} agree, one can build a local coordinate system such that the entire metric is continuous across $\S$ too. In that case, one can define a unique unit normal $n^\mu$, which we choose to point from $V^-$ towards $V^+$, as shown.}
\label{fig:glued}
\end{figure}

The space-time is given by an $(n+1)$-dimensional Lorentzian manifold $(V,g)$.
Let us consider the case where $(V,g)$ possesses two different regions, say with different matter contents or different gravitational fields, separated by a border. This border will locally be a hypersurface $\S\subset V$ which can have any causal character, the physically more interesting case arising when it is {\em timelike}, which we will assume throughout in this paper. $\Sigma$ divides the manifold $V$ into two regions $V^\pm$, as shown schematically in Fig.\ref{fig:glued}. 

The metrics $g_{\mu\nu}^\pm$ are assumed to be smooth on $V^\pm$ respectively ($\mu,\nu,\dots  = 0,1,\dots,n$). An important observation is that one can actually deal with two different coordinate systems on $V^\pm$. In fact, this is {\em needed} in most practical problems, as one is usually given two distinct solutions of the field equations that are to be matched: for instance, one solution describing an interior with matter and another describing vacuum; or a background solution upon which a localized perturbation, such as a wave front or a shell of matter, propagates. Thus, we will be presented with two sets of local coordinates $\{x^\mu_{\pm}\}$ {\em with no relation whatsoever}, each valid on the corresponding part $V^\pm$ \cite{I}. 

Two corresponding timelike hypersurfaces $\Sigma^{\pm}\subset V^{\pm}$ which bound the regions $V^{\pm}$ must be chosen on each $\pm$-side to be matched. Of course, these two hypersurfaces are to be identified in the final glued spacetime, so that they must be diffeomorphic. The junction of $V^+$ with $V^-$ by identifying $\S^+$ with $\S^-$ 
depends crucially on the particular diffeomorphism used for this identification, hence we assume that this has already
been chosen and is known. The glued global manifold $V$ is defined as the disjoint union of $V^{+}$ and $V^{-}$ with diffeomorphically related points of $\Sigma^{+}$ and $\Sigma^{-}$ identified. This unique
hypersurface is the {\em matching hypersurface} we denote simply by $\Sigma$.

Let $\{\xi^a\}$ be a set of local coordinates on $\Sigma$ ($a,b,\dots =1,\dots , n$). Then, there are two parametric representations 
$$
x_{\pm}^\mu =x_{\pm}^\mu (\xi^a)
$$ 
of $\Sigma$, one for each imbedding into each of $V^\pm$. As explained in the Appendix \ref{App:B} , in order to have well defined curvature tensors in the sense of distributions we need a global metric which is at least continuous across $\S$. As is known  \cite{CD,MS}, this happens if and only if the two first fundamental forms $h^{\pm}$ of $\Sigma$ inherited from both sides $V^\pm$ agree. This agreement requires the equalities on $\Sigma$
\be
h^+_{ab} = h_{ab}^- ,\hspace{1cm}
h^\pm_{ab} \equiv g_{\mu\nu}^\pm (x(\xi))\frac{\partial x_\pm^\mu}{\partial \xi^a}\frac{\partial x_\pm^\nu}{\partial \xi^b} 
\label{h=h}
\ee
and implies that one can build local coordinate systems in which the metric can be extended to be continuous across $\Sigma$. 
The unique metric defined on the entire manifold that coincides with
$g^{\pm}$ in the respective $V^{\pm}$ and is continuous across $\S$ is denoted simply by $g$.

Let $n^\pm_\mu$ be the unit normals to $\Sigma$ as seen from $V^\pm$ respectively. They are fixed up to a sign by the conditions
$$
n^\pm_\mu \frac{\partial x_\pm^\mu}{\partial \xi^a}=0, \quad n^\pm_\mu n^{\pm\mu}=1
$$
and one must choose one of them (say $n^-_\mu$) pointing outwards from $V^-$ and the other ($n^+_\mu$) pointing towards $V^+$. Hence, the two bases on the tangent spaces at any point of $\S$
$$
\{n^{+\mu},\frac{\partial x_+^\mu}{\partial \xi^a}\} \quad \quad \leftrightarrow \quad \quad\{n^{-\mu},\frac{\partial x_-^\mu}{\partial \xi^a}\}
$$ 
agree and are then identified, so we drop the $\pm$ (even though, in explicit calculations, one can still use both versions using the two coordinate systems on each side). We denote by $\vec{e}_a$ the vector fields tangent to $\S$ defined by the above imbeddings
$$
\vec{e}_a \defi \left.\frac{\partial x_+^\mu}{\partial \xi^a}\frac{\partial}{\partial x^\mu_+}\right|_\S =\left.\frac{\partial x_-^\mu}{\partial \xi^a}\frac{\partial}{\partial x^\mu_-}\right|_\S .
$$
Note that $\{\vec{e}_a\}$ are defined only on $\S$. The basis dual to $\{n^\mu,e^\mu_a\}$ is denoted by
$$
\{n_\mu, \omega^a_\mu\}
$$
where the one-forms $\bm{\omega}^a$ are characterized by
$$
n^\mu \omega^a_\mu =0, \hspace{1cm} e^\mu_b \omega^a_\mu =\delta^a_b .
$$
The space-time version of the first fundamental form, which is now unique due to (\ref{h=h}), is given by the projector to $\Sigma$ (defined only on $\S$)
\be
h_{\mu\nu}=g_{\mu\nu}-n_\mu n_\nu .\label{proj}
\ee
Notice that 
$$
n^\mu h_{\mu\nu} =0, \hspace{1cm} h_{\mu\rho} h^\rho{}_\nu =h_{\mu\nu}, \hspace{1cm} h^\mu{}_\mu =n, \hspace{1cm}  h_{\mu\nu}e^\mu_a e^\nu_b =h_{ab}
$$
and that 
$$
e^\mu_a =h_{ab}\, \omega^b_\nu \, g^{\nu\mu} , \hspace{1cm} e^\mu_c \omega^c_\nu =h^\mu_\nu \, .
$$

Despite all the above, the extrinsic curvatures, or second fundamental forms, inherited by $\S$ from both sides $V^\pm$ will be, in principle, different, because the derivatives of the metric are not continuous in general. We denote them by $K^\pm_{\mu\nu}$, and they are defined, as usual, by 
$$
K^\pm_{\mu\nu}\defi h^\rho{}_{\nu}h^\sigma_\mu \nabla^\pm_\rho n_\sigma \hspace{1cm} K^\pm_{\mu\nu}=K^\pm_{\nu\mu}
$$
where only tangent derivatives are involved. Obviously $n^\mu K^\pm_{\mu\nu}=0$, thus only the $n(n+1)/2$ components  tangent to $\Sigma$ are non-identically vanishing.
In terms of the imbeddings these components are given by
\be
K^\pm_{ab} \equiv -n^\pm_\mu \left(\frac{\partial^2 x^\mu_\pm}{\partial\xi^a\partial\xi^b}+\Gamma^{\pm \mu}_{\rho\sigma}\frac{\partial x_\pm^\rho}{\partial \xi^a} \frac{\partial x_\pm^\sigma}{\partial \xi^b}\right), \label{2FF}
\ee
which is adapted to explicit calculations. These components correspond to the second fundamental form, defined as a tensor
in $\S$ by
$$
K^\pm_{ab} =-n_\mu e^\rho_a\nabla^\pm_\rho e^\mu_b =e^\mu_b  e^\rho_a\nabla^\pm_\rho n_\mu \, .
$$

As shown in the Appendix \ref{App:C}, the Riemann tensor can be computed in the distributional sense and acquires, in general, a singular part proportional to the distribution $\deltaS$ supported on $\S$ ---which is defined in Appendix \ref{App:B}---:
\be
\underline{R}^\alpha{}_{\beta\mu\nu}=R^{+\alpha}{}_{\beta\mu\nu}\otheta + R^{-\alpha}{}_{\beta\mu\nu}(\1-\underline{\theta})+\deltaS H^\alpha{}_{\beta\mu\nu} .\label{Riedist}
\ee
Here
$H^\alpha{}_{\beta\mu\nu}$ is called the singular part of the Riemann tensor distribution and as shown in the Appendix \ref{App:C}
reads
$$
H^\a{}_{\beta\lambda\mu}= n_{\lambda}\left
[\Gamma^{\alpha}_{\beta\mu} \right ] - n_{\mu} \left [ 
\Gamma^{\alpha}_{\beta\lambda} \right ]
$$
where the square brackets always denote the jump of the enclosed object across $\S$ according to the definition (\ref{discont}) given in Appendix \ref{App:B}. 
We can  provide a more interesting formula for this singular part. First, note that from the general formula for the discontinuities of derivatives (\ref{discf}) in Appendix \ref{App:D2} we have
$$
\left[\partial_\a g_{\mu\nu} \right]= n_\a \zeta_{\mu\nu}
$$
for some symmetric tensor field $\zeta_{\mu\nu}$ defined only on $\S$. This immediately gives
\be
 \left [ \Gamma^{\alpha}_{\beta\lambda} \right ] =\frac{1}{2} \left(\zeta^\a{}_\b n_\l +\zeta^\a{}_\l n_\b -n^\a \zeta_{\b\l} \right)\label{Gammadisc}
\ee
which implies
\be
H_{\alpha\beta\lambda\mu} = \frac{1}{2}\left(- n_{\alpha}\zeta_{\beta\mu}n_{\lambda}+n_{\alpha}\zeta_{\beta\lambda}n_{\mu}-n_{\beta}\zeta_{\alpha\lambda}n_{\mu}+n_{\beta}\zeta_{\alpha\mu}n_{\lambda} \right).\label{HRie0}
\ee
Note that this expression is invariant under the change $\zeta_{\mu\nu}\longrightarrow \zeta_{\mu\nu}+n_\mu X_\nu +n_\nu X_\mu$ for arbitrary $X_\mu$ and thus only the part of $\zeta_{\mu\nu}$ {\em tangent} to $\S$ enters into the formula. Actually, one can prove the existence of $C^1$, piecewise $C^\infty$, changes of coordinates that remove any normal part of $\zeta_{\m\n}$ arising in (\ref{Gammadisc}) ---see, e.g., \cite{MS}. Thus, from now on we assume that such a change has been performed and we will restrict ourselves to assuming that $\zeta_{\m\n}$ is tangent to $\S$: $n^\m \zeta_{\m\n} =0$. But using (\ref{2FF}) together with (\ref{Gammadisc}) we deduce
\be
K^+_{ab} -K^-_{ab} = -n_\mu \left[\Gamma^\mu_{\rho\sigma}\right] e^\rho_a e^\sigma_b= \frac{1}{2} \zeta_{\rho\sigma}e^\rho_a e^\sigma_b \label{discK}
\ee
that is to say, the tangent part of $\zeta_{\mu\nu}$ is characterized by the difference of the two $\pm$-second fundamental forms. Thus, defining the {\em jump}  on $\Sigma$ of the second fundamental form as usual 
\be
      \left[K_{\mu\nu}\right]\defi K^{+}_{\mu\nu}-K^{-}_{\mu\nu}, \hspace{1cm} n^{\mu}\left[K_{\mu\nu}\right] =0 \label{Kdisc}
\ee
we can rewrite (\ref{HRie0}) as the desired formula for the singular part of the Riemann tensor distribution:
\be
H_{\alpha\beta\lambda\mu} = - n_{\alpha}\left[K_{\beta\mu}\right]n_{\lambda}+n_{\alpha}\left[K_{\beta\lambda}\right]n_{\mu}-n_{\beta}\left[K_{\alpha\lambda}\right]n_{\mu}+n_{\beta}\left[K_{\alpha\mu}\right]n_{\lambda} .\label{HRie}
\ee
This important formula informs us that {\em the singular part of the Riemann tensor distribution vanishes if, and only if, the jump of the second fundamental form vanishes}. 

By contractions on (\ref{Riedist}) we get (with obvious notations):
\begin{itemize}
\item The Ricci tensor distribution
\be
\underline{R}_{\b\m}=R^+_{\b\m}\otheta +R^-_{\b\m} (\1-\otheta) +H_{\b\m} \deltaS \label{Ricdist}
\ee
where its singular part is given by
\be
H_{\beta\mu}\defi H^\rho{}_{\beta\rho\mu} =-\left[K_{\beta\mu}\right] -\left[K^\rho{}_{\rho}\right] n_\beta n_\mu .\label{HRic}
\ee
Thus, {\em the singular part of the Ricci tensor distribution vanishes if, and only if, the jump of the second fundamental form vanishes, hence, if and only if that of the full Riemann tensor distribution does}. 
\item The scalar curvature distribution
\be
\underline R = R^+\otheta +R^- (\1-\otheta) + H \deltaS \label{scalardist}
\ee
whose singular part reads
\be
H\defi H^\rho{}_{\rho}=-2\left[K^\mu{}_{\mu}\right]  . \label{Hscalar}
\ee
It follows that {\em the singular part of the scalar curvature distribution vanishes if, and only if, the jump of the \underline{trace} of second fundamental form vanishes}. 
\item And the Einstein tensor distribution
\be
\underline{G}_{\b\m} \defi \underline{R}_{\b\m}-\frac{1}{2} g_{\b\m} \underline R =G^+_{\b\m}\otheta +G^-_{\b\m} (\1-\otheta) +{\cal G}_{\b\m} \deltaS \label{Gdist}
\ee
with a singular part
\be
{\cal G}_{\beta\mu} = -\left[K_{\beta\mu}\right]+h_{\beta\mu}\left[K^\rho{}_{\rho}\right] , \hspace{1cm} n^\mu {\cal G}_{\beta\mu} =0 \label{HG}
\ee
which is tangent to $\S$.
\end{itemize}

A general result proven in \cite{MS} is that the second Bianchi identity holds in the distributional sense:
$$
\nabla_{\rho}\underline{R}^\alpha{}_{\beta\mu\nu}+\nabla_{\mu}\underline{R}^\alpha{}_{\beta\nu\rho}+\nabla_{\nu}\underline{R}^\alpha{}_{\beta\rho\mu}=0
$$ 
from where one deduces by contraction
$$
\nabla^\beta \underline{G}_{\beta\mu}=0
$$
for the Einstein tensor distribution. By using (\ref{Gdist}) and the general formula (\ref{nablaT1}) this implies
\be
0= \nabla^\beta \underline{G}_{\beta\mu} = n^\beta \left[G_{\beta\mu}\right] \underline{\delta}^\Sigma +\nabla^\beta \left({\cal G}_{\beta\mu}\underline{\delta}^\Sigma \right)\, .\label{divG=0}
\ee
The second summand on the righthand side is computed according to the general formula (\ref{nablaTdelta}) in Appendix \ref{App:D1}
\bean
\nabla^\beta \left({\cal G}_{\beta\mu}\underline{\delta}^\Sigma \right)=
g^{\beta\rho}\nabla_\rho \left({\cal G}_{\beta\mu}\underline{\delta}^\Sigma \right)=
g^{\beta\rho}\N_\sigma \left({\cal G}_{\b\m} n_\r n^\sigma\deltaS \right)+g^{\b\r} 
h^\l_\r\nabla_\l {\cal G}_{\b\m} \deltaS =h^{\r\l}
\nabla_\l {\cal G}_{\r\m}\,  \deltaS 
\eean
which, via (\ref{nabla=nabla1}) finally gives
$$
\nabla^\beta \left({\cal G}_{\beta\mu}\underline{\delta}^\Sigma \right)=\left(\overline\nabla^\b {\cal G}_{\b\m}-K^\S_{\r\sigma}{\cal G}^{\r\sigma}n_\m \right)\deltaS \, .
$$
Introducing this into (\ref{divG=0}) we arrive at
$$
0=\underline{\delta}^\Sigma\left(n^\beta \left[G_{\beta\mu}\right] +\overline\nabla^\beta {\cal G}_{\beta\mu}-\frac{1}{2}n_\mu {\cal G}^{\rho\sigma}(K^+_{\rho\sigma}+K^-_{\rho\sigma})\right)
$$
which implies, by taking the normal and tangent components, the following relations 
\bea
(K^+_{\rho\sigma}+K^-_{\rho\sigma}){\cal G}^{\rho\sigma} = 2n^\beta n^\mu \left[ G_{\beta\mu}\right]=2n^\beta n^\mu \left[ R_{\beta\mu}\right]-[R], \label{1}\\
\overline\nabla^\beta {\cal G}_{\beta\mu}=-n^\rho h^\sigma{}_\mu \left[ G_{\rho\sigma}\right]=-n^\rho h^\sigma{}_\mu \left[ R_{\rho\sigma}\right] \label{2} .
\eea
(These equations can also be obtained \cite{I} by using part of the Gauss and Codazzi equations for $\Sigma$ on both sides, specifically (\ref{gauss}) and ({\ref{coda}) in Appendix \ref{App:D1}).

{\bf Remark:}  A very important remark is that all formulae in this section are {\em purely geometric}, independent of any field equations, and therefore valid in any theory of gravity based on a Lorentzian manifold.

\section{Quadratic gravity}
\label{sec:quadratic_grav}
We are going to concentrate on the case of quadratic theories of gravity because, apart from its own intrinsic interest and as we are going to discuss, they allow for cases where gravitational double layers arise.
Let us consider a quadratic theory of gravity in $n+1$ dimensions described by the Lagrangian density
\begin{equation}
\mathcal{L} = \frac{1}{2\kappa}\left(R-2\Lambda + \au R^2 + \ad R_{\m\n}R^{\m \n} + \at R_{\a \b \m \n} R^{\a \b \m \n}\right)+\mathcal{L}_{matter},\label{lag}
\end{equation}
where $\kappa =c^4/8\pi G$ is the gravitational coupling constant, $\Lambda$ is the cosmological constant, $\au,\ad,\at$ are three constants selecting the particular theory, and $\mathcal{L}_{matter}$ is the Lagrangian density describing the matter fields. $\Lambda^{-1}$ and $\au,\ad,\at$ have physical units of $L^{2}$. 
The field equations derived from this Lagrangian read (see e.g. \cite{F} and references therein)
\be
G_{\alpha\beta}+\Lambda g_{\a\b}+G^{(2)}_{\alpha\beta}=\kappa T_{\alpha\beta}, \label{fe}
\ee
where $T_{\alpha\beta}$ is the energy-momentum tensor of the matter fields derived from $\mathcal{L}_{matter}$,
$G_{\alpha\beta}$ is the Einstein tensor and $G^{(2)}_{\alpha\beta}$ encodes the part that comes from the quadratic terms:
\begin{eqnarray}
G^{(2)}_{\a \b}&=& 2\left\{\frac{}{} \au R R_{\a \b} -2\at R_{\a \m}R_{\b}^\m + \at R_{\a \r \m \n}R_\b{}^{\;\r \m \n} + (\ad + 2\at)R_{\a \m \b \n}R^{\m \n} \right.\nonumber\\
&&\left. - \left(\au + \frac{1}{2}\ad + \at\right)\nabla_\a \nabla_\b R+\left( \frac{1}{2}\ad + 2 \at \right)\Box R_{\a \b}\right\}\nonumber\\
&&-\frac{1}{2} g_{\a \b}\left\{(\au R^2 + \ad R_{\m\n}R^{\m \n} + \at R_{\r \g \m \n} R^{\r \g \m \n}) - (4\au + \ad) \Box R \right\} \label{eq:G2}
\end{eqnarray}
where $\Box\defi g^{\m\n}\nabla_\m\nabla_\n$ is notation for the D'Alembertian in $(V,g)$. 

If we want to find the proper junction conditions, or a description of thin shells or braneworlds in these theories, we have to resort to the distributional calculus (see Appendices) and use the formulas provided in the previous section. 
Then, in order to have the Lagrangian density as well as the tensor $G^{(2)}_{\a \b}$ well defined in a distributional sense ---so that the field equations (\ref{fe}) are sensible mathematically---, one has to avoid any multiplication of singular distributions (such as ``$\deltaS \deltaS$''). One could also hope for some cancellation of such terms between different parts of the Lagrangian, and of $G^{(2)}_{\a \b}$, and this is discussed in the following subsection for completeness, but one has to bear in mind that these cancellations are probably ill defined anyway, and thus not relevant. In order to properly deal with products of distributions
we would need a more general calculus, based e.g. on Colombeau algebras \cite{Colombeau, Vickers}, and hope that those cancellations certainly occur and are well defined.

\subsection{Dubious possible cancellation of non-linear $\deltaS \deltaS$ terms}
Let us start by examining the Lagrangian (\ref{lag}) recalling that the different curvature terms possess now singular parts proportional to $\deltaS$, as given in
(\ref{HRie}) and its contractions (\ref{HRic}) and (\ref{Hscalar}). One could naively compute the products of these singular parts arising from the quadratic terms in (\ref{lag}) and collect them in a common-factor fashion. The result would be a term of type
$$
\deltaS\deltaS \left( 2\kappa_1 [K_\r^\r]^2 +2 \kappa_2 [K_{\a\b}][K^{\a\b}] \right)
$$
where we have introduced the abbreviations
\begin{equation}
\kappa_1\defi 2\au+\ad/2,\qquad
\kappa_2\defi 2\at+\ad/2 .
\label{def:kappas}
\end{equation}
to be used repeatedly in what follows. Then, one should require the vanishing of the term in brackets. A similar naive compilation should be performed with the non-linear distributions arising from the quadratic terms in the field equations (\ref{eq:G2}). Imposing again that the full combination must vanish, and separating the resulting condition into its normal and tangent parts to $\Sigma$ we would find
\begin{eqnarray}
&&\left \lbrace \kappa_1 [K_\r^\r]^2 + \kappa_2 (3[K^{\mu \nu}][K_{\mu \nu}] - 2[K_\r^\r]^2 ) \right \rbrace n_\a n_\b \label{eq:normaldeltadelta}\\
&&+\kappa_1 [K_\r^\r](2[K_{\a\b}] - [K_\r^\r]h_{\a\b}) + \kappa_2 (2[K_\r^\r][K_{\a\b}]-[K_{\mu\nu}][K^{\mu\nu}]h_{\a\b}) = 0.\label{eq:tangentdeltadelta}
\end{eqnarray}
The normal (\ref{eq:normaldeltadelta}) and tangent (\ref{eq:tangentdeltadelta}) parts should vanish separately. In particular the trace of the tangent part reads
\begin{equation}
\kappa_1 [K_\r^\r]^2(2 - n) + \kappa_2 (2[K_\r^\r]^2-n[K_{\mu\nu}][K^{\mu\nu}]) = 0.\label{eq:tracetangentdeltadelta}
\end{equation}
We see directly that $\kappa_1 = \kappa_2 = 0$ solves (\ref{eq:normaldeltadelta}) and (\ref{eq:tangentdeltadelta}), but in order to find all solutions we compute the determinant of the system (\ref{eq:normaldeltadelta}) and (\ref{eq:tracetangentdeltadelta}). This yields
\begin{eqnarray}
 (3-n)[K_\r^\r]^2([K_\r^\r]^2  - [K^{\mu \nu}][K_{\mu \nu}]) = 0.
\end{eqnarray}
Take first $[K_\r^\r]=0$. Then, (\ref{eq:normaldeltadelta}) and (\ref{eq:tracetangentdeltadelta}) reduce to $\kappa_2 [K_{\mu  \nu}][K^{\mu \nu}] = 0$. If $[K_\r^\r]\neq 0$ but
$[K_\r^\r]^2  = [K^{\mu \nu}][K_{\mu \nu}]$, (\ref{eq:normaldeltadelta}) reads  $(\kappa_1 + \kappa_2)[K_\r^\r]^2=0$ and (\ref{eq:tracetangentdeltadelta}) is redundant since it becomes $(\kappa_1 + \kappa_2)[K_\r^\r]^2(2[K_{\a\b}] - [K_\r^\r] h_{\a\b})=0$. Thus, $\kappa_1 + \kappa_2 =0$ would follow.
Finally, if $n=3$ (and $[K_\r^\r]^2 \neq 0$), (\ref{eq:normaldeltadelta}),
(\ref{eq:tangentdeltadelta}) and (\ref{eq:tracetangentdeltadelta}) yield
a new possibility not considered so far, summarized in
\begin{equation}
[K_{\a\b}] = \frac{1}{3} h_{\a\b} \Rightarrow [K_\r^\r] = 1, \quad [K_{\a\b}][K^{\a\b}] = \frac{1}{3}, \quad \quad \kappa_1 -\kappa_2=0.
\end{equation}

In short, each of the following possibilities would seem to allow for the mutual annihilation of "$\delta^\Sigma \delta^\Sigma$" terms in (\ref{eq:G2}) ---and in (\ref{lag})---:
\begin{enumerate}
\item $\kappa_1 = \kappa_2 = 0$.
\item $[K_\r^\r] = 0$ and $\kappa_2 = 0$.
\item $[K_\r^\r]^2 = [K_{\m\n}][K^{\m\n}] = 0$.
\item $[K_\r^\r]^2 = [K_{\m\n}][K^{\m\n}] \neq 0$ and $\kappa_1 + \kappa_2 = 0$.
\item If the spacetime is 4-dimensional, $\kappa_1 - \kappa_2 = 0$ and $[K_{\a\b}] = h_{\a\b}/3$.
\end{enumerate}

Despite we have included this analysis here for completeness, we should not forget that these cases are not mathematically correct, and therefore they should not be taken seriously unless a more rigorous study is performed showing its feasibility. To understand the problems behind these naive calculations, we want to emphasize that there is no known way to give a sensible meaning to $\deltaS \deltaS$, let alone to things such as $f \deltaS \deltaS$. Thus, taking for granted that combinations of type $f_1\deltaS \deltaS + f_2 \deltaS\deltaS$ are related to $(f_1+f_2)\deltaS \deltaS$ is, at least, dubious. Such difficulties were, for instance, noted in \cite{DD} for the Gauss-Bonnet case ---corresponding to the possibility 1 above---, and one has to resort to analyzing thick shells, that is, layers with a finite width, or to a setting more general than distributions,
such as the theory of nonlinear generalized functions described in \cite{Colombeau, Vickers} and references therein. The thin shell formalism is simply not available. 
Therefore, we will abandon this route for now, and in this paper we will concentrate on the generic and well-defined cases analyzed in the next subsection.

\subsection{Well defined possibilities: no $\deltaS\deltaS$ terms}
The only mathematically well-defined possibilities in the available theory of distributions for the thin shell formalism, as just argued, are those where no $\deltaS\deltaS$ term ever arises, leading to two different possibilities if we let aside the case of GR (defined by $\au =\ad =\at =0$):
\begin{enumerate}
\item If either $\ad$ or $\at$ is different from zero, then products of the Ricci tensor by itself, or by the Riemann tensor, appear in (\ref{eq:G2}) and these are ill-defined if the singular parts (\ref{HRie}) and (\ref{HRic}) are non-zero. Thus, we must demand that the singular parts (\ref{HRie}) and (\ref{HRic}) vanish which happens, as proven above, if and only if the jump of the second fundamental form vanishes. Thus, in this situation it is indispensable to require
\be
\left[K_{\m\n}\right] =0 .\label{Kdisc=0}
\ee
In this case, all the curvature tensors are tensor distributions associated to tensor fields ---see Appendix \ref{App:A}---with possible discontinuities across $\S$. Observe that then the Lagrangian density (\ref{lag})  is also a well defined, locally integrable, function.
\item If on the other hand $\ad=\at=0$, then only products of $R$ by itself or by the Ricci tensor appear in (\ref{eq:G2}), and thus it is enough to demand that $R$ is a locally integrable function without singular part. Hence, in this case it is enough to require that (\ref{Hscalar}) vanishes, that is to say, that the trace of the second fundamental form has no jump: $[K^\r_\r]=0$. Observe that, again, the Lagrangian density (\ref{lag}) is in this case a well-defined locally integrable function.
\end{enumerate}

In any of the above two possibilities, expression (\ref{fe}) with (\ref{eq:G2}) has a remarkable property: {\em there are no terms quadratic in derivatives of the curvature tensors}. Taking into account that tensor distributions can be covariantly differentiated according to the rules explained in the appendices, the derivatives of the curvature tensors may have singular parts and still the field equations (\ref{fe}) are mathematically sound. This opens the door for the existence of matching hypersurfaces which represent {\em double layers}. Case 2 above was extensively treated in \cite{Senovilla13,Senovilla14,Senovilla15}, where gravitational double layers were found for the first time. Therefore, we will here concentrate in the more general case 1, and thus we will assume hereafter that (\ref{Kdisc=0}) holds. Notice that (\ref{Kdisc=0}) coincide precisely with the matching conditions that are needed in General Relativity to avoid distributional matter contents, as follows from (\ref{HG}) together with the Einstein field equations. 

Once (\ref{Kdisc=0}) is enforced, the lefthand side of the field equations (\ref{fe}) can be computed in the distributional sense. From (\ref{Riedist}) and (\ref{Kdisc=0}) we know that the Riemann tensor distribution 
\[
\underline R_{\alpha\beta\mu\nu}=R^+_{\alpha\beta\mu\nu}\underline{\theta}+R^-_{\alpha\beta\mu\nu}(\1-\underline{\theta}),
\]
is actually associated to a locally integrable (and piecewise differentiable) tensor field. However, this tensor field may be discontinuous across $\S$, and thus $[R_{\a\b\m\n}]$ may be non-vanishing. This leads, when computing covariant derivatives of $\underline R_{\alpha\beta\mu\nu}$, to singular terms proportional $\deltaS$ and its derivatives. And these are going to arise in $\underline G^{(2)}_{\a\b}$. 
Thus, the energy-momentum tensor on the righthand side of (\ref{fe}) must be treated as a tensor distribution and contain such terms, localized on $\S$, giving the energy-matter contents of the thin shell or double layer. 

In order to compute this matter content supported on $\S$ we only have to calculate the singular part of $\underline G^{(2)}_{\a\b}$, because ${\cal G}_{\a\b}$ in (\ref{Gdist}) vanishes  as follows from (\ref{Kdisc=0}) with (\ref{HG}). But
the only terms in (\ref{eq:G2})  that are relevant for this singular part
are $\nabla_\alpha\nabla_\beta R$ and $\Box R_{\alpha\beta}$ (and its contraction $\Box R$). More precisely,
we need to obtain the singular part of the expression
\begin{eqnarray}
&&  - \left(2\au + \ad + 2\at\right)\N_\alpha \N_\beta \underline R+\left( \ad + 4 \at \right)\Box \underline R_{\a \b}+ \left(2\au +\frac{1}{2} \ad\right) \Box \underline R\, g_{\a\b}\nonumber\\
&&=- \left(\kappa_1+\kappa_2\right)\N_\alpha \N_\beta \underline R+2\kappa_2\Box \underline R_{\a \b}+ \kappa_1 \Box \underline R\, g_{\a\b} .
\label{eq:sing_expression}
\end{eqnarray}  
This is the purpose of the next section.

\section{Energy-momentum on the layer $\S$}
\label{sec:compute_deltas}
From (\ref{Ricdist}) and the assumption (\ref{Kdisc=0}) we know that
$$
\underline R_{\a \b} = R_{\a \b}^+ \, \otheta + R_{\a \b}^- (\1-\otheta)
$$
from where, using the general formula (\ref{nablaT1}) twice we deduce
\begin{eqnarray}
\N_\n \underline R_{\a \b} &=& \N_\n R_{\a \b}^+ \, \otheta + \N_\n R_{\a \b}^- (\1-\otheta) + [R_{\a \b}] n_\n \deltaS,\nonumber \\
\N_\m \N_\n \underline R_{\a \b} &=& \N_\m \N_\n R_{\a \b}^+\,  \otheta + \N_\m \N_\n R_{\a \b}^- (\1-\otheta) + [\N_\n R_{\a \b}]n_\mu \deltaS + \N_\mu \left([R_{\a \b}]n_\n \deltaS \right). \label{eq:dd_ricci_dist}
\end{eqnarray}
Via contractions here, or directly from (\ref{scalardist}), we also obtain
\begin{eqnarray}
\underline R &=& R^+ \otheta + R^- (\1-\otheta),\nonumber \\
\N_\n \underline R &=& \N_\n R^+ \otheta + \N_\n R^- (\1-\otheta) + [R]n_\n \deltaS, \nonumber \\
\N_\m \N_\n \underline R &=& \N_\m \N_\n R^+ \otheta + \N_\m \N_\n R^- (\1-\otheta) + [\N_\n R]n_\mu \deltaS + \N_\mu \left([R]n_\n \deltaS \right)
\label{eq:2_der_R}
\end{eqnarray}
as well as 
\bea
\Box \underline R_{\alpha\beta}= \Box R_{\a \b}^+ \otheta + \Box R_{\a \b}^- (\1-\otheta) + n^\r[\nabla_\r R_{\a\b}] \deltaS + g^{\m \n} \N_\mu \left([R_{\a \b}]n_\n \deltaS\right), \\
\Box \underline R = \Box R^+ \otheta + \Box R^- (\1-\otheta) + n^\r[\nabla_\r R] \deltaS + g^{\m \n} \N_\mu \left([R]n_\n \deltaS\right). 
\eea

Thus, we need to control the discontinuities of the Ricci tensor and the scalar curvature, and also to provide an expression for the singular distribution $\N_\mu \left([R_{\a \b}]n_\n \deltaS \right)$ supported on $\S$. The general formula (\ref{nablaTdelta}) provides
$$
 \N_\mu\left(n_\nu [R_{\a\b}]\deltaS\right)=\N_\r \left(  [R_{\a\b}] n_\mu n_\nu n^\r \deltaS \right)+ \left\{ h^\r_\m \nabla_\r (n_\n [R_{\a\b}] )-K^{\r}{}_\r \, [R_{\a\b}] n_\m n_\n \right\}  \deltaS .
$$
At this point we introduce a 4-covariant tensor distribution $\underline{\Delta}_{\m\n\a\b}$ with support on $\Sigma$, which takes care of the first summand here and is defined by
$$
\underline{\Delta}_{\m\n\a\b}\defi \nabla_\r \left([R_{\a\b}]n_\m n_\n n^\r \, \deltaS \right) 
$$
or equivalently by
$$
\left \langle \underline{\Delta}_{\m\n\a\b}, Y^{\m \n \a \b}\right \rangle \defi - \int_\Sigma [R_{\a \b}] n_\n n_\m n^\r \N_\r Y^{\m \n \a \b} d \sigma .
$$
Note that $\underline{\Delta}_{\m\n\a\b}=\underline{\Delta}_{\n\m\a\b}=\underline{\Delta}_{\m\n\b\a}$. In summary, we have
$$
 \N_\mu\left(n_\nu [R_{\a\b}]\deltaS\right)= \underline{\Delta}_{\m\n\a\b}+  \left\{n_\n h^\r_\m \nabla_\r [R_{\a\b}] +[R_{\a\b}] (K_{\m\n}-K^{\r}{}_\r \, n_\m n_\n)  \right\} \deltaS
$$
and therefore (\ref{eq:dd_ricci_dist}) becomes
\bean
\N_\m \N_\n \underline R_{\a \b} = \N_\m \N_\n R_{\a \b}^+\,  \otheta + \N_\m \N_\n R_{\a \b}^- (\1-\otheta) +\underline{\Delta}_{\m\n\a\b}\\
+\left\{ [\N_\n R_{\a \b}]n_\mu+ n_\n h^\r_\m \nabla_\r [R_{\a\b}] +[R_{\a\b}] (K_{\m\n}-K^{\r}{}_\r \, n_\m n_\n)  \right\} \deltaS .
\eean
From the general formula (\ref{disc1f}), conveniently generalised,  we have
\begin{equation}
  \label{eq:[d_ricci]}
  [\nabla_\rho R_{\beta\mu}]=n_\rho r_{\beta\mu}+h^\sigma_\rho\nabla_\sigma[R_{\beta\mu}],
\end{equation}
where 
\be
r_{\beta\mu}\defi n^\r[\nabla_\r R_{\b\m}], \hspace{1cm} r_{\b\m} =r_{\m\b} \label{rab}
\ee 
are the discontinuities of the normal derivatives of the Ricci tensor. Thus, we finally get
\bea
\N_\m \N_\n \underline R_{\a \b} = \N_\m \N_\n R_{\a \b}^+\,  \otheta + \N_\m \N_\n R_{\a \b}^- (\underline{1}-\otheta) +\underline{\Delta}_{\m\n\a\b}\hspace{1cm} \nonumber \\
+\left\{ r_{\a\b} \, n_\n n_\mu+  n_\m h^\r_\n \nabla_\r [R_{\a\b}]+n_\n h^\r_\m \nabla_\r [R_{\a\b}] +[R_{\a\b}] (K_{\m\n}-K^{\r}{}_\r \, n_\m n_\n)  \right\} \deltaS .\label{nablanablaRic}
\eea
Observe that the entire singular part is symmetric in $(\a\b)$ and in $(\m\n)$.

From (\ref{nablanablaRic}) we immediately get all the sought terms. First, by contracting with $g^{\a\b}$ we find \cite{Senovilla13,Senovilla14,Senovilla15}
\bea
\N_\m \N_\n \underline R = \N_\m \N_\n R^+\,  \otheta + \N_\m \N_\n R^- (\underline{1}-\otheta) +\underline{\Delta}_{\m\n}\hspace{1cm} \nonumber \\
+\left\{ b n_\n n_\mu+  n_\m \Nb_\n [R]+n_\n \Nb_\m [R] +[R] (K_{\m\n}-K^{\r}{}_\r \, n_\m n_\n)  \right\} \deltaS \label{nablanablaR}
\eea
where \cite{Senovilla14,Senovilla15}
\be
b\defi r^\r_\r =n^\r\nabla_\r [R] \label{b}
\ee
measures the discontinuity on the normal derivative of the scalar curvature, and \cite{Senovilla14}
$$
\underline{\Delta}_{\mu\nu} \defi g^{\a\b}\underline{\Delta}_{\m\n\a\b}
$$
is a 2-covariant symmetric tensor distribution with support on $\S$ acting as follows\footnote{There are some errata in the formulae for $\underline{\Delta}_{\m\n}$ and $\underline{\Omega}_{\m\n}$ in \cite{Senovilla13}, and for $\underline{t}_{\m\n}$ in \cite{Senovilla14,Senovilla15}:  in all cases $Y$ must be replaced by $Y^{\m\n}$.}
\begin{equation}
\left \langle \dl_{\m \n}, Y^{\m \n}\right \rangle \defi - \int_\Sigma [R] n_\n n_\m n^\r \N_\r Y^{\m \n} d \sigma ; \hspace{1cm} \dl_{\m \n} = \nabla_\r \left([R] n_\m n_\n n^\r \deltaS \right).
\label{def:dl}
\end{equation}
Similarly, contracting (\ref{nablanablaRic}) with $g^{\m\n}$ we readily get
\begin{eqnarray}
\Box R_{\a \b} = \Box R_{\a \b}^+ \otheta + \Box R_{\a \b}^- (\underline{1}-\otheta) + r_{\a \b} \deltaS + g^{\m \n} \underline{\Delta}_{\m\n\a\b}
\label{eq:box_ricci_dist}
\end{eqnarray}
where the last distribution acts as follows
\bean
  \left\langle g^{\mu\nu}\underline{\Delta}_{\m\n\a\b}, Y^{\alpha\beta}\right\rangle=
  \left\langle \underline{\Delta}_{\m\n\a\b},  g^{\mu\nu}Y^{\alpha\beta}\right\rangle=
  -\int_\Sigma[R_{\alpha\beta}]n_\nu n_\mu n^\r\N_\r(Y^{\alpha\beta}g^{\mu\nu})d\sigma\nonumber\\
   =-\int_\Sigma[R_{\alpha\beta}] n^\r\N_\r Y^{\alpha\beta}d\sigma ; \hspace{2cm} g^{\mu\nu}\underline{\Delta}_{\m \n \a \b}=\N_\r\left([R_{\alpha\beta}]n^\r\deltaS\right). 
\eean
Finally, by tracing either of (\ref{nablanablaR}) or (\ref{eq:box_ricci_dist}) we easily derive
\begin{equation}
  \Box \underline{R} = \Box R^+ \otheta + \Box R^- (\underline{1}-\otheta) + b\,  \deltaS +  \dl,
\end{equation}
where we have introduced the notation $\dl\defi g^{\m \n} \dl_{\m \n}$. Note that \cite{Senovilla13}
\[
\left\langle \dl, Y\right\rangle=\left\langle g^{\mu\nu}\dl_{\m \n}, Y\right\rangle=-\int_\Sigma [R]n^\r \N_\r Y d\sigma ;
\hspace{1cm} \dl = \N_\r \left([R] n^\r \deltaS\right)
\]

What we have proven is that the distribution $\underline{G}^{(2)}_{\a\b}$ takes the following form
\be
\underline{G}^{(2)}_{\a\b} = G^{(2)+}_{\a\b}\otheta + G^{(2)-}_{\a\b}(\1-\otheta)+\widetilde{G}_{\a\b} \deltaS +\mathscr{G}_{\a\b} \label{G2dist}
\ee
where 
\be
\widetilde{G}_{\a\b}=2\kappa_2 r_{\a\b}+\kappa_1 b g_{\a\b} -(\kappa_1 +\kappa_2) \left\{ b n_\a n_\b+  n_\a \Nb_\b [R]+n_\b \Nb_\a [R] +[R] (K_{\a\b}-K^{\r}{}_\r \, n_\a n_\b)  \right\} ,\label{tildeG}
\ee
and after a trivial rearrangement 
\begin{eqnarray}
\mathscr{G}_{\alpha\beta}=\kappa_1 \left( g_{\a\b}\dl - \dl_{\a \b}\right) + \kappa_2 \left( 2 g^{\m \n} \underline{\Delta}_{\m \n \a \b}  - \dl_{\a \b}\right) . \label{arranged1}
\end{eqnarray}
From (\ref{arranged1}) we define two new 2-covariant tensor distributions with support on $\S$ \cite{Senovilla14}:
\be
\underline{\Omega}_{\a\b} \defi g_{\a\b}\dl - \dl_{\a \b}=\N_\r\left( [R] h_{\a\b} n^\r\deltaS\right) ; \hspace{7mm} \left<\underline{\Omega}_{\a\b},Y^{\a\b} \right>=
-\int_\S [R] h_{\a\b}\,  n^\r\nabla_\r Y^{\a\b}d\sigma  \label{Omega}
\ee
and
\be
\underline{\Phi}_{\a\b} \defi g^{\m \n} \underline{\Delta}_{\m \n \a \b}  -\frac{1}{2} \dl_{\a \b}-\frac{1}{2}\underline{\Omega}_{\a\b}
=\N_\r \left( [G_{\a\b}] n^\r \deltaS\right)  ; \hspace{3mm} 
\left<\underline{\Phi}_{\a\b},Y^{\a\b} \right>=
-\int_\S [G_{\a\b}]\,  n^\r\nabla_\r Y^{\a\b} d\sigma \label{Phi }
\ee
(recall that $[G_{\a\b}]$ is tangent to $\S$, $n^\a[G_{\a\b}]=0$,  due to (\ref{1}) and (\ref{2}) together with the vanishing of ${\cal G}_{\a\b}$ as follows from (\ref{HG}) and (\ref{Kdisc=0})). With these definitions, (\ref{arranged1}) is rewritten simply as
\be
\mathscr{G}_{\a\b} = (\kappa_1+\kappa_2)  \underline{\Omega}_{\a\b} +2\kappa_2 \underline{\Phi}_{\a\b}; \hspace{1cm} \mathscr{G}_{\a\b} = \N_\r \left(\left\{(\kappa_1+\kappa_2) [R] h_{\a\b} +2\kappa_2[G_{\a\b}]\right\}n^\r \deltaS \right). \label{Glayer}
\ee

Given the structure (\ref{G2dist}), the field equations (\ref{fe}) can only be satisfied if the energy-momentum tensor on the righthand side is a tensor distribution with the following terms
\be
\underline T_{\mu\nu}=T^+_{\mu\nu} \underline\theta +T^-_{\mu\nu} (\1-\underline\theta)+\widetilde{T}_{\m\n} \deltaS + \underline{t}_{\mu\nu} \label{emt0}
\ee
where $\widetilde{T}_{\m\n}$ is a symmetric tensor field defined only on $\S$ and $\underline{t}_{\m\n}$ is by definition the singular part of $\underline{T}_{\m\n}$ with support on $\S$ {\em not proportional} to $\deltaS$. We perform an orthogonal decomposition of $\tilde{T}_{\m\n}$ into tangent, normal-tangent and normal parts with respect to $\S$
\be
\widetilde{T}_{\m\n} =\tau_{\mu\nu}+\tau_\mu n_\nu +\tau_\nu n_\mu +\tau n_\mu n_\nu \label{emtortog}
\ee
with
\bean
\tau_{\m\n} \defi h^\r_\m h^\sigma_\n \widetilde{T}_{\r\sigma}, \hspace{3mm} \tau_{\m\n} =\tau_{\n\m} , \hspace{3mm} n^\m\tau_{\m\n}=0; \hspace{7mm} \tau_{\m} \defi h^\r_\m \widetilde{T}_{\r\n}n^\n, \hspace{3mm} n^\m \tau_\m =0; \hspace{7mm} \tau \defi n^\m n^\n \widetilde{T}_{\m\n}
\eean
so that
\be
\underline T_{\mu\nu}=T^+_{\mu\nu} \underline\theta +T^-_{\mu\nu} (\underline{1}-\underline\theta)+\left(\tau_{\mu\nu}+\tau_\mu n_\nu +\tau_\nu n_\mu +\tau n_\mu n_\nu \right) \underline\delta^{\Sigma} + \underline{t}_{\mu\nu}.  \label{emt}
\ee

Following \cite{Senovilla14,Senovilla15} the proposed names for the objects in (\ref{emt}) supported on $\S$, with their respective explicit expressions, are:
\begin{enumerate}
\item the {\em energy-momentum tensor} $\tau_{\a\b}$ on $\S$, given by
\be
\kappa \tau_{\a\b} =-(\kappa_1 + \kappa_2) [R] K_{\a\b}+\kappa_1 b h_{\a\b} + 2 \kappa_2 r_{\m\n} h^\m_\a h^\n_\b .\label{tauexc}
\ee
$\tau_{\a\b}$ is the only quantity usually defined in standard shells.
\item the {\em external flux momentum}  $\tau_{\a}$ defined by
\be
\kappa \tau_\a =-(\kappa_1 + \kappa_2)\overline{\N}_\a [R] + 2\kappa_2 r_{\m\n}n^\m h^\n_\a . \label{tauex}
\ee
This momentum vector describes normal-tangent components of $\underline{T}_{\mu\nu}$ supported on $\S$. Nothing like that exists in GR.
\item the {\em external pressure or tension} $\tau$ 
\be 
\kappa \tau = (\kappa_1 + \kappa_2) [R] K^\r_\r  + \kappa_2 (2r_{\m\n}n^\m n^\n -b), \label{taue} 
\ee
Taking the trace of (\ref{tauexc}) one obtains a relation between $b$, $\tau$ and the trace of $\tau_{\mu\nu}$:
\be
\kappa \left(\tau^{\rho}{}_{\rho}+\tau\right) = (\kappa_1 n+\kappa_2) b \label{para_sacar_b}
\ee
The scalar $\tau$ measures the total normal pressure/tension supported on $\S$. Again, such a scalar does not exist in GR.
\item the {\em double-layer energy-momentum tensor distribution} $\underline t_{\a\b}$, which is defined by
\be
\kappa \underline t_{\a\b} = \mathscr{G}_{\a\b} =  \N_\r \left(\left\{(\kappa_1+\kappa_2) [R] h_{\a\b} +2\kappa_2[G_{\a\b}]\right\}n^\r \deltaS \right)
\ee
or, equivalently, by acting on any test tensor field $Y^{\a\b}$ as
\be
\kappa \left<\underline t_{\a\b},Y^{\a\b}\right> = - \int_\Sigma \left\{(\kappa_1+\kappa_2) [R] h_{\a\b} +2\kappa_2[G_{\a\b}]\right\} n^\rho\nabla_\rho Y^{\a\b} \, . \label{t}
\ee
$\underline{t}_{\a\b}$ is a symmetric tensor distribution of ``delta-prime'' type: it has support on $\S$ but its product with objects intrinsic to $\S$ is not defined unless their extensions off $\S$ are known. As argued in \cite{Senovilla14,Senovilla15}, $\underline{t}_{\a\b}$ resembles the energy-momentum content of double-layer surface charge distributions, or ``dipole distributions'', with strength 
\be
\kappa \mu_{\a\b}\defi (\kappa_1+\kappa_2) [R] h_{\a\b} +2\kappa_2[G_{\a\b}] ,\hspace{7mm} \mu_{\a\b}=\mu_{\b\a}, \hspace{5mm} n^\a \mu_{\a\b} =0. \label{strength}
\ee
We note in passing that 
\be
\kappa \mu^\r{}_\r = (\kappa_1 n +\kappa_2) [R], \hspace{1cm} \kappa \underline{t}^\r{}_\r =(\kappa_1 n +\kappa_2)\underline\Delta 
\label{tracestrength}
\ee
The appearance of such double layers is remarkable, as ``massive dipoles" do not exist.  However, in quadratic theories of gravity they arise, as we have just shown, in the generic situation when thin shells are considered. In this case, $\underline{t}_{\a\b}$ seems to represent the idealization of abrupt changes, or jumps, in the curvature of the space-time.
\end{enumerate}

\section{Curvature discontinuities}\label{sec:nG2}
In the next section, we are going to derive the field equations satisfied by the energy-momentum quantities (\ref{tauexc}), (\ref{tauex}), (\ref{taue}) and (\ref{strength}) supported on  $\S$. To that end, we have to perform a detailed calculation of the discontinuities of the field equations (\ref{fe}): they obviously include the discontinuities of the energy-momentum tensor $T_{\m\n}$ which must be related to the energy-momentum content concentrated on $\S$.

The discontinuity of the lefthand side of (\ref{fe}) contains $[G_{\a\b}^{(2)}]$ (actually, we will only need $n^\a[G_{\a\b}^{(2)}]$) and this involves discontinuities of quadratic terms in the Riemann tensor, such as
$[R^2]$, $[R_{\alpha\beta}R^{\alpha\beta}]$, $[R_{\alpha\beta\mu\nu}R^{\alpha\beta\mu\nu}]$,
$[R R_{\a \b}]$, $[R_{\a \m}R_{\b}^\m]$, $[R_{\a \r \m \n}R_\b{}^{\;\r \m \n}]$
and $[R_{\a \m \b \n}R^{\m \n}]$, as well as discontinuities of derivatives of the curvature tensors, such as $[\nabla_\a\nabla_\b R]$, $[\Box R_{\a\b}]$ or $[\Box R]$. Thus, we have to use systematically the rules (\ref{discprod}) and either of (\ref{disc1f}) or (\ref{disc1f1}) supplemented with (\ref{Kdisc=0}), and we also need to have some knowledge on the discontinuities of the Riemann tensor (and its derivatives). 

\subsection{Discontinuities of the curvature tensors}
Thus, let us start by controlling the allowed discontinuities of the Riemann tensor across $\S$. From the requirement (\ref{Kdisc=0}) we know that $\zeta_{\m\n}=0$ and thus
$$
[\Gamma^\a_{\b\m}] =0.
$$
Then, the general formula (\ref{discf}) gives
$$
\left[\partial_\l \Gamma^\a_{\b\m} \right]= n_\l \gamma^\a{}_{\b\m} 
$$
for some functions $\gamma^\a{}_{\b\m}$ such that $\gamma^\a{}_{\b\m}=\gamma^\a{}_{\m\b}$, and therefore
\be
\left[R_{\a\b\l\m}\right]=n_\l \gamma_{\a\b\m} -n_\m \gamma_{\a\b\l} .
\ee
The antisymmetry of $R_{\a\b\l\m}$ in $[\a\b]$ implies then that $\gamma_{(\a\b)\m}=-B_{\a\b} n_\m$ for some symmetric tensor $B_{\a\b}=B_{\b\a}$ defined only on $\S$. Hence
$$
\gamma_{\a\b\m} =\tilde{\gamma}_{\a\b\m} -B_{\a\b} n_\m, \hspace{1cm} \tilde{\gamma}_{\a\b\m}=-\tilde{\gamma}_{\b\a\m}
$$
and
$$
\left[R_{\a\b\l\m}\right]=n_\l \tilde\gamma_{\a\b\m} -n_\m \tilde\gamma_{\a\b\l} .
$$
However, the symmetry of $ \gamma_{\a\b\m}$ in $(\b\m)$ implies $\tilde\gamma_{\a[\b\m]}-B_{\a[\b} n_{\m]}=0$ as well as $\tilde\gamma_{[\a\b\m]}=0$ from where one easily derives
$$
\tilde\gamma_{\a\b\m}=2\tilde\gamma_{\m[\b\a]}= n_\a B_{\b\m} -n_\b B_{\a\m}
$$
and we recover the standard formula \cite{MS} 
\begin{equation}
  \label{eq:[Rie]}
  [R_{\alpha\beta\lambda\mu}]=n_\alpha n_\lambda B_{\beta\mu}-n_\lambda n_\beta B_{\alpha\mu}-n_\mu n_\alpha B_{\beta\lambda}+n_\mu n_\beta B_{\alpha\lambda}.
\end{equation}
As argued after formula (\ref{HRie}), there is no loss of generality by assuming that $B_{\alpha\beta}$ is tangent to $\Sigma$, i.e. such that 
$$
B_{\alpha\beta}n^\alpha=0 \hspace{1cm} \Longrightarrow \hspace{3mm} B_{\a\b} = B_{ab}\omega^a_\a \omega^b_\b .
$$
Given this plus the symmetry of $B_{\a\b}$ there are $(n^2+n)/2$ independent allowed discontinuities for the curvature tensor, all encoded in $B_{ab}$.

Successive contractions on (\ref{eq:[Rie]}) provide
\begin{equation}
  \label{eq:[ricci]}
  [R_{\beta\mu}]=B_{\beta\mu}+\frac{1}{2} n_\beta n_\mu [R],\qquad [R]=2B^\r_\r,
\end{equation}
or equivalently
\be
B_{\b\m} =   [R_{\beta\mu}]- \frac{1}{2}[R] n_\beta n_\mu =[G_{\b\m}] +\frac{1}{2} h_{\b\m} [R] \, .\label{B=G}
\ee
In other words, the $n(n+1)/2$ allowed independent discontinuities of the Riemann tensor can be chosen to be the discontinuities of the $\S$-tangent part of the Einstein tensor (or equivalently, of the Ricci tensor). 

\subsection{Discontinuities of the curvature tensors derivative}
Concerning the covariant derivative of the Riemann tensor, the general formula (\ref{disc1f}) leads to
\begin{equation}
  \label{eq:[d_Rie]}
  [\nabla_{\rho}R_{\alpha\beta\lambda\mu}]=n_\rho r_{\alpha\beta\lambda\mu}+ h^{\sigma}_\rho\nabla_\sigma[R_{\alpha\beta\lambda\mu}],
\end{equation}
where $r_{\alpha\beta\mu\nu}$ is a tensor field defined only on $\S$ and with the symmetries of a Riemann
tensor. Using the second Bianchi identity for the Riemann tensor the previous formula implies
$$
n_{[\rho} r_{\alpha\beta]\lambda\mu}+h_{\sigma[\rho} \nabla^\sigma[R_{\alpha\beta]\lambda\mu}]=0
$$
which, on using (\ref{eq:[Rie]}) and after some calculations, implies the following structure for 
$r_{\alpha\beta\mu\nu}$: 
\begin{eqnarray}
r_{\alpha\beta\m\n}&=& K_{\alpha\mu}B_{\nu\beta}
-K_{\alpha\nu}B_{\mu\beta}+K_{\beta\nu}B_{\mu\alpha}-K_{\beta\mu}B_{\nu\alpha}\nonumber\\
&+&\left(\Nb_\mu B_{\rho\nu}-\Nb_\nu B_{\rho\mu}\right)(n_\alpha h^\rho_\beta-n_\beta h^\rho_\alpha)
+\left(\Nb_\a B_{\rho\b}-\Nb_\b B_{\rho\a}\right)(n_\m h^\rho_\n-n_\n h^\rho_\m)\nonumber\\
&+& n_\a n_\m \r_{\b\n} -n_\a n_\n \r_{\b\m} - n_\b n_\m \r_{\a\n}+ n_\b n_\n \r_{\a\m}, \label{eq:rel_tB}
\end{eqnarray}
where $\r_{\b\m}$ is a new symmetric tensor field, defined only on $\S$ and tangent to $\S$, $n^\b\r_{\b\m}=0$, which encodes the {\em allowed new} independent discontinuities of the covariant derivative of the Riemann tensor. There are $n(n+1)/2$ of those again.  
As far as we know, relation (\ref{eq:rel_tB}) has only been derived in \cite{MS}. 

Contraction of (\ref{eq:rel_tB}) leads to the equation (\ref{eq:[d_ricci]}), but now with an explicit expression for the discontinuity of the normal derivative of the Ricci tensor which reads, on using (\ref{B=G})
\begin{eqnarray}
r_{\b\n} &=&\r_{\b\n} + K^\r_\r B_{\b\n}  +\frac{1}{2}[R] K_{\b\n} -K_{\r\b}B^\r_\n- B_{\r\b}K^\r_\n \nonumber\\
&-&n_\b\Nb_\r [G^\r _\n] -n_\n \Nb_\r [G^\r _\b] \nonumber\\
&+&n_\b n_\n \r^\a_\a, \label{disnormDRicc}
\end{eqnarray}
where a natural orthogonal decomposition of $r_{\b\m}$ appears: the first line is its complete tangent part which, given that $\r_{\b\n}$ entails the allowed new independent discontinuities, is in itself a symmetric tensor field tangent to $\S$ codifying those discontinuities. We are going to denote it by
\be
{\cal R}_{\b\m}\defi h^\r_\b h^\sigma_\m r_{\rho\sigma} =h^\r_\b h^\sigma_\m n^\l [\N_\l R_{\rho\sigma}]; \label{rtangent}
\ee
the second line is its tangent-normal part, which is completely determined by the covariant derivative within $\S$ of the discontinuity of the Einstein tensor
\begin{equation}
n^\b h^\n_\m  r_{\b\n} = -\Nb^\r [G_{\r\m}];
  \label{eq:dRnnt}
\end{equation}
and finally, the third line gives the total normal component of $r_{\b\m}$, which can be related to the discontinuity (\ref{b}) of the normal derivative of $R$ by simply taking the trace $r^\r_\r =b $ leading to 
\be
r_{\b\m}n^\b n^\m =\frac{b}{2}+K^{\r\sigma} [G_{\r\sigma}] . \label{eq:dRnnt1}
\ee
Using this we get a useful relation for the trace of $\mathcal{R}_{\a\b}$, that does not depend on $\rho_{\a\b}$
\be
\mathcal{R}_\a^\a = \frac{b}{2} - K^{\rho \sigma}[G_{\rho \sigma}]. \label{traceRcal}
\ee
\subsection{Second-order derivative discontinuities}
Let us now consider the jumps in the second derivatives of the Ricci tensor. The starting point is equation (\ref{eq:[d_ricci]}). We can find an expression for the second summand there by differentiating (\ref{eq:[ricci]}) along
$\Sigma$ and using the general rule (\ref{nabla=nabla1}) (see Appendix \ref{App:D1}),
\begin{equation}
  \label{eq:[d_diff_ricci]}
  h^\sigma_\rho\nabla_\sigma[R_{\beta\mu}]=\frac{1}{2}n_\beta n_\mu\Nb_\rho [R]
  +n_{(\mu}\left(K_{\beta)\rho}[R] -2B_{\beta)\lambda}K^\lambda_\rho\right) +\Nb_\rho B_{\beta\mu}.
\end{equation}
The jumps of the second-order derivatives of the Ricci tensor,
due to the general formula (\ref{disc1f}), can be written as 
\begin{equation}
  \label{eq:[dd_ricci]}
 [\nabla_\lambda\nabla_\rho R_{\beta\mu}]=n_\lambda A_{\rho\beta\mu}+h^\sigma_\lambda\N_\sigma [\N_\rho R_{\beta\mu}]
\end{equation}
where $A_{\rho\beta\mu}=A_{\rho(\beta\mu)}$ is a shorthand for
$$
A_{\rho\beta\mu}=n^\l [\N_\l \nabla_\rho R_{\beta\mu}].
$$ 
The last term
$h^\sigma_\lambda\N_\sigma [\N_\rho R_{\beta\mu}]$ can be further expanded
by first using (\ref{eq:[d_ricci]}) to obtain
\[
h^\sigma_\lambda\N_\sigma [\N_\rho R_{\beta\mu}]=
K_{\lambda\rho}r_{\beta\mu} +n_\rho h^\sigma_\lambda\N_\sigma r_{\beta\mu}
+h^\sigma_\lambda \N_\sigma \left(h^\gamma_\rho \N_\gamma[R_{\beta\mu}]\right).
\]
and then computing the last summand here, which leads to
\begin{eqnarray}
  &&h^\sigma_\lambda\N_\sigma [\N_\rho R_{\beta\mu}]=
  K_{\lambda\rho} r_{\beta\mu}+ 2 n_{(\mu} K_{\beta)(\lambda}\Nb_{\rho)}[R] +[R] K_{\rho(\beta}K_{\mu)\lambda}-4 K^\gamma_{(\rho}\Nb_{\lambda)} B_{\gamma(\beta} n_{\mu)}\nonumber \\
  &&\qquad   +n_\beta n_\mu\left(\frac{1}{2}\Nb_\lambda\Nb_\rho [R]
     -[R]K_\lambda^\sigma K_{\rho\sigma}
     +2K^\gamma_\rho K^\sigma_\lambda B_{\sigma\gamma}\right)
       \nonumber\\
  &&\qquad + \left(\Nb_\lambda K^\gamma_\rho-n_\rho K^\sigma_\lambda K^\gamma_\sigma\right)\left([R] h_{\gamma(\beta}-2B_{\gamma(\beta}\right)n_{\mu)}
     -\frac{1}{2} n_\mu n_\beta n_\rho K_\lambda^\sigma\Nb_\sigma [R]
     \nonumber\\
  &&\qquad +\Nb_\lambda \Nb_\rho B_{\beta\mu} -2K^\gamma_\rho B_{\gamma(\beta}K_{\mu)\lambda}
     - n_\rho K^\sigma_\lambda \Nb_\sigma B_{\beta\mu}
     + n_\rho h_\lambda^\sigma \N_\sigma r_{\beta\mu}.
\label{eq:A}
\end{eqnarray}
Let us stress the fact that all the terms in
the first two lines in the above expression are symmetric in $(\lambda \rho)$.

Concerning $A_{\rho\beta\mu}$,
let us first decompose it into normal and tangential parts
by
$$
A_{\rho\beta\mu}=n_\rho A_{\beta\mu}+h^\gamma_\rho A_{\gamma\beta\mu},
\hspace{1cm} A_{\b\m} \defi n^\r A_{\rho\beta\mu}, \hspace{5mm} A_{\b\m} =A_{\m\b} .
$$
In order to obtain an expression for $h^\gamma_\rho A_{\gamma\beta\mu}$
we take the antisymmetric part of (\ref{eq:[dd_ricci]}) with respect to $[\lambda \rho]$,
and contract with $n^\lambda$.
For the left hand side of (\ref{eq:[dd_ricci]}) we 
use the Ricci identity applied to the Ricci tensor at both sides $V^\pm$, and take the
difference of the limits on $\S$, so that
\[
[(\nabla_{\lambda}\nabla_{\rho} -\nabla_{\r}\nabla_{\l})R_{\beta\mu}]=[R^\gamma{}_{\beta\rho\lambda}R_{\gamma\mu}]+[R^\gamma{}_{\mu\rho\lambda}R_{\beta\gamma}]. 
\]
For the right hand side of (\ref{eq:[dd_ricci]}), after the contraction with $n^\lambda$, we get
\[
A_{\rho\beta\mu}-n^\lambda n_{\r} A_{\l\beta\mu}-n^\lambda h^\sigma_\r \N_\sigma [\N_\l R_{\beta\mu}]=
h^\gamma_{\r} A_{\gamma\beta\mu}-n^\lambda h^\sigma_\r \N_\sigma [\N_\l R_{\beta\mu}].
\]
Isolating $h^\gamma_\rho A_{\gamma\beta\mu}$ and using (\ref{eq:A}) for the last term of the above equation,
it is then straightforward to obtain
\begin{eqnarray}
  &&A_{\rho\beta\mu}=n_\rho A_{\beta\mu}
  +n^\nu[R^\gamma{}_{\beta\rho\nu}R_{\gamma\mu}]
  +n^\nu[R^\gamma{}_{\mu\rho\nu}R_{\beta\gamma}]
  +h^\sigma_\rho\N_\sigma r_{\beta\mu}\nonumber \\
  &&\qquad -\frac{1}{2}n_\mu n_\beta K^\sigma_\rho\Nb_\sigma [R]
  -K^\sigma_\rho \Nb_\sigma B_{\beta\mu}
  -[R] K^\sigma_\rho K_{\sigma(\beta} n_{\mu)}
   + 2K^\sigma_\rho K^\gamma_\sigma B_{\gamma(\beta} n_{\mu)}.
  \label{eq:ddRn}
\end{eqnarray}
The expression for $[\nabla_\lambda\nabla_\rho R_{\beta\mu}]$ now follows
by combining (\ref{eq:[dd_ricci]}) with
(\ref{eq:A}) and (\ref{eq:ddRn}). After little rearrangements, that reads
\begin{eqnarray}
  &&[\nabla_\lambda\nabla_\rho R_{\beta\mu}]=n_\lambda n_\rho A_{\beta\mu}
  +n_\lambda n^\nu\left([R^\gamma{}_{\beta\rho\nu}R_{\gamma\mu}]+[R^\gamma{}_{\mu\rho\nu}R_{\beta\gamma}]\right)
  +2n_{(\lambda} h_{\rho)}{}^\sigma\N_\sigma r_{\beta\mu}\nonumber\\
  &&\qquad - n_\mu n_\beta n_{(\lambda} K_{\rho)}{}^\sigma\Nb_\sigma [R]
  -2 n_{(\lambda} K_{\rho)}{}^\sigma \Nb_\sigma B_{\beta\mu}
  - 2 [R] n_{(\lambda}  K_{\rho)}{}^\sigma K_{\sigma(\beta} n_{\mu)}\nonumber \\
  &&\qquad +4n_{(\lambda} K_{\rho)}{}^\sigma K^\gamma_\sigma B_{\gamma(\beta} n_{\mu)}
  + 2 n_{(\mu} K_{\beta)(\lambda}\Nb_{\rho)}[R] +[R] K_{\rho(\beta}K_{\mu)\lambda}-4 K^\gamma{}_{(\rho}\Nb_{\lambda)} B_{\gamma(\beta} n_{\mu)}\nonumber \\
  &&\qquad   +n_\beta n_\mu\left(\frac{1}{2}\Nb_\lambda\Nb_\rho [R]
     -[R] K_\lambda^\sigma K_{\rho\sigma}
     +2K^\gamma_\rho K^\sigma_\lambda B_{\sigma\gamma}\right) +  K_{\lambda\rho} r_{\beta\mu}
       \nonumber\\
  &&\qquad + \Nb_\lambda K^\gamma_\rho\left([R] h_{\gamma(\beta}-2B_{\gamma(\beta}\right)n_{\mu)}
          +\Nb_\lambda \Nb_\rho B_{\beta\mu}-2K^\gamma_\rho B_{\gamma(\beta}K_{\mu)\lambda}.
       \label{eq:[ddRicci]}
\end{eqnarray}
We must stress the fact that there are still terms in (\ref{eq:[ddRicci]}), i.e. $A_{\beta\mu}$ and $r_{\beta\mu}$,
that are not completely independent.

The contraction of (\ref{eq:[ddRicci]}) with $g^{\rho\lambda}$ yields
\begin{eqnarray}
  &&[\Box R_{\beta\mu}]=A_{\beta\mu}+K r_{\beta\mu}+
     n_\mu n_\beta\left(\frac{1}{2} \Box [R]-[R]K^{\r\sigma} K_{\r\sigma}+2K^{\sigma\rho}K_\rho^\gamma B_{\sigma\gamma}\right) \nonumber \\
  &&\qquad +2n_{(\mu} h^\lambda_{\beta)}\left(\Nb_\rho [R] K^\rho_\lambda-2K^{\gamma\rho}\Nb_\rho B_{\gamma\lambda}+\frac{1}{2}\Nb_\r K^\r_\l - \Nb_\rho K^{\rho\gamma}B_{\gamma\lambda} \right)\nonumber\\
  &&\qquad +[R] K_{\rho\beta}K_{\mu}^\rho+\Boxb B_{\beta\mu}-2K^\gamma_\rho K^\rho_{(\mu}B_{\beta)\gamma},
\label{eq:[box_ricci]}
\end{eqnarray}
while contracting with $g^{\beta\mu}$ we obtain \cite{Senovilla13}
\begin{equation}
  [\N_\nu\N_\mu R]=A^\r_\r n_\nu n_\mu+2n_{(\nu}\Nb_{\mu)}b
-2n_{(\nu} K_{\mu)}^\lambda\Nb_\lambda [R]+b K_{\nu\mu}+\Nb_\nu\Nb_\mu [R].
  \label{eq:[dd_ricciS]}
\end{equation}
From any of the previous we readily have
\begin{equation}
  [\Box R]=A^\r_\r+bK+\Boxb [R] .
\label{eq:[box_ricciS]}
\end{equation}

The energy-momentum quantities (\ref{tauexc}-\ref{taue}) will arise from the discontinuities of the {\em normal} components of the lefthand side of (\ref{fe}). In other words, we will only need to consider $n^\a[G^{(2)}_{\a\b}]$. 
Observe then that $A_{\beta\mu}$ only appears in $[\Box R_{\beta\mu}]$,
and since we only need the terms contracted with the normal once, in particular $n^\beta[\Box R_{\beta\mu}]$,
we are only interested in controlling $n^\beta A_{\beta\mu}$. This can be done by using the identities $2\N^\r R^\pm_{\r\m}=\N_\m R^\pm$ at both sides of $\S$, and taking the difference after one further differentiation:
\begin{equation}
  n^\nu[\N_\nu\N^\rho R_{\rho\mu}]=\frac{1}{2}n^\nu [\N_\nu\N_\mu R].
\label{eq:bianchi_contr}
\end{equation}
The lefthand side here comes from (\ref{eq:[dd_ricci]}) combined with (\ref{eq:ddRn}) after one contraction,
whereas for the righthand side 
we simply have to contract (\ref{eq:[dd_ricciS]}) with $n^\nu$.
Equation (\ref{eq:bianchi_contr}) is thus found to be equivalent to
\begin{eqnarray}
 n^\rho A_{\rho\mu}
  +n^\sigma\left(-[R^\gamma_\sigma R_{\gamma\mu}]+[R_{\gamma\mu\rho\sigma} R^{\gamma\rho}]\right)+h^{\beta\sigma}\N_\sigma r_{\beta\mu}-K^{\beta\sigma}\Nb_\sigma B_{\beta\mu}\nonumber \\
  -n_\mu\left(\frac{1}{2}[R] K_{\r\sigma} K^{\r\sigma} -K^{\sigma\beta}K_\sigma^\gamma B_{\gamma\beta}\right)
  =\frac{1}{2}\left(A^\r_\r n_\mu+ \Nb_\mu b -K_\mu^\lambda\Nb_\lambda [R]\right).
\label{eq:for_nddRn}
\end{eqnarray}

\subsection{Discontinuities of terms quadratic in the curvature}
Now, let us concern ourselves with the many terms in (\ref{eq:G2}) quadratic in curvature tensors. To start with,
using (\ref{eq:[ricci]}) with (\ref{discprod}) we readily obtain
\begin{eqnarray}
  &&[R_{\alpha\beta}R^{\alpha\beta}]=2 [R^{\alpha\beta}]\RS_{\alpha\beta} = \left(B^{\a\b}+\frac{1}{2} [R] n^\a n^\b\right)\RS_{\alpha\beta}, \\
  &&[R R_{\alpha\beta}]=\RS [R_{\a\b}] +[R] \RS_{\a\b} = \RS\left(B_{\alpha\beta}+\frac{1}{2}n_\alpha n_\beta [R]\right) +{R} \RS_{\alpha\beta}, \label{eq:[RR]}\\
   &&    [R^2]=2[R]\RS.
\label{eq:[RR]1}
\end{eqnarray}

Regarding $n^\alpha[R_{\a \m}R_{\b}^\m]$, let us first consider the contraction $n^\sigma n^\mu [R^\gamma_\sigma R_{\gamma\mu}]$. The chain of equalities
\begin{equation}
  n^\sigma n^\mu [R^\gamma_\sigma R_{\gamma\mu}]=2n^\sigma n^\mu\RS{}^\gamma_\sigma[R_{\gamma\mu}]=[R] n^\gamma n^\mu\RS{}_{\gamma\mu}
  \label{eq:nn[RR]1}
\end{equation}
follows from  (\ref{eq:[ricci]}) and (\ref{discprod}). Half-adding the two $\pm$ equations (\ref{gauss})  and using the result in (\ref{eq:nn[RR]1}) we derive
\begin{equation}
  n^\sigma n^\mu [R^\gamma_\sigma R_{\gamma\mu}]=\frac{1}{2} [R] (\RS-\Rb+(K^\rho_\rho)^2 -  K_{\r\sigma} K^{\r\sigma} ).
  \label{eq:nn[RR]}
\end{equation}
Analogous procedures using the Gauss equation (\ref{gauss0}) accordingly yield
\begin{eqnarray}
  &&n^\sigma h^\mu_\n [R^\gamma_\sigma R_{\gamma\mu}]=B_{\a\n}(\Nb_\b K^{\b\a}-\Nb^\a K^\r_\r )+\frac{1}{2}[R](\Nb_\a K^\a_\n-\Nb_\n K^\r_\r ),\label{eq:ne[RR]}\\
  &&n^\alpha n^\beta[R_{\alpha\mu\beta\nu} R^{\mu\nu}]=
\left(2\RS_{\mu\nu}- \Rb_{\mu\nu} +K^\r_\r  K_{\mu\nu}-K_{\mu\sigma}K^\sigma_\nu\right) B^{\mu\nu},
\label{eq:nn[RieRic]}\\
  &&n^\alpha h^\beta_\l[R_{\alpha\mu\beta\nu} R^{\mu\nu}]=
  (\Nb_\b K_{\l\a}-\Nb_\l K_{\b\a}) B^{\a\b}-(\Nb_\a K^{\a\b}-\Nb^\b K^\r_\r )B_{\b\l}.
\label{eq:ne[RieRic]}\\
  &&n^\alpha n^\beta[R_{\alpha\rho\mu\nu} R_\beta{}^{\rho\mu\nu}]=4n^\alpha n^\beta[R_{\alpha\mu\beta\nu} R^{\mu\nu}]-4\RS_{\r\sigma} B^{\r\sigma},
\label{eq:nn[RieRie]}\\
  &&n^\alpha h^\beta_\l [R_{\alpha\rho\mu\nu} R_\beta{}^{\rho\mu\nu}]=2 B^{\a\b}(\Nb_\a K_{\l\b}-\Nb_\l K_{\a\b}),\\
  &&[R_{\alpha\beta\mu\nu} R^{\alpha\beta\mu\nu}]=2n^\alpha n^\beta[R_{\alpha\rho\mu\nu} R_\beta{}^{\rho\mu\nu}]= 8B^{\a\b}(\RS_{\a\b}-\Rb_{\a\b}+K^\r_\r  K_{\a\b}-K_{\a\r}K^\r_\b).
     \label{eq:[RieRie]}
\end{eqnarray}

\subsection{Discontinuities of the quadratic part $[G^{(2)}_{\a\b}]$}

We are now ready to compute the full $n^\a[G^{(2)}_{\a\b}]$. To keep track of the different terms, we split the compilation of terms in three parts, corresponding to the terms multiplied by either of the three constants $a_1, a_2,a_3$ in (\ref{eq:G2}). 
\begin{itemize}
\item Terms with $\au$:

The terms in (\ref{eq:G2}) that go with $\au$ are
\[
G^{(2)a_1}_{\alpha\beta}\defi 2 R R_{\alpha\beta}-2\N_\beta\N_\alpha R
-\frac{1}{2}g_{\alpha\beta}R^2+2g_{\alpha\beta}\Box R,
\]
and we can compute their jump using (\ref{eq:[RR]}), (\ref{eq:[dd_ricciS]}) and (\ref{eq:[box_ricciS]}) to obtain
\begin{equation}
  n^\alpha n^\beta[G^{(2)a_1}_{\alpha\beta}]=2[R] \RS_{\alpha\beta}n^\alpha n^\beta
  +2b K^\r_\r+2\Boxb [R]
\label{eq:nnGa1}
\end{equation}
and
\begin{equation}
  n^\alpha h^\beta_\m [G^{(2)a_1}_{\alpha\beta}]=
  2[R] \RS_{\alpha\beta}n^\alpha h^\beta_\m-2\Nb_\m b+2 K_\m^\a \Nb_\a [R].
\label{eq:neGa1}
\end{equation}

\item Terms with $\ad$:

The terms in (\ref{eq:G2}) relative to $\ad$ are 
\[
G^{(2)a_2}_{\alpha\beta}\defi 2 R_{\alpha\mu\beta\nu} R^{\mu\nu}-\N_\beta\N_\alpha R
+\Box R_{\alpha\beta}-\frac{1}{2}g_{\alpha\beta}\left(R_{\m\n} R^{\m\n}-\Box R\right).
\]
Before using (\ref{eq:nn[RieRic]}) and (\ref{eq:ne[RieRic]}) it is convenient to
write down $n^\alpha [\Box R_{\alpha\beta}]$ using (\ref{eq:[box_ricci]}) combined with (\ref{eq:for_nddRn}),
since some terms simplify. With the help of
(\ref{eq:nn[RieRic]}), (\ref{eq:ne[RieRic]}), (\ref{eq:[dd_ricciS]}), (\ref{eq:[RR]})
and (\ref{rtangent}-\ref{eq:dRnnt1}) it is then  easy to get
\begin{eqnarray}
  n^\alpha n^\beta[G^{(2)a_2}_{\alpha\beta}] &=& \frac{b}{2} K^\r_\r + \Boxb [R] + \frac{1}{4}[R](\RS-\Rb + (K^\r_\r)^2)-\frac{3}{4}[R] K_{\r\sigma} K^{\r\sigma}
                                                 \nonumber\\
                                     &&+ {\cal R}_{\r\sigma} K^{\r\sigma} + \Nb_\r\Nb_\m [G^{\r\m}]   +B^{\m\n}(\RS_{\m\n} - \Rb_{\m\n} + K^\r_\r K_{\m\n}) , \\
 n^\alpha h^\beta_\m[G^{(2)a_2}_{\alpha\beta}]&=& -\frac{1}{2}\Nb_\m b + \frac{3}{2} K^\a_\m \overline{\N}_\a [R]+ [R] \left( \overline{\N}_\a K^\a_\m - \frac{1}{2} \overline{\N}_\m K\right) -\Nb_\a {\cal R}^\a_\m + K^\a_\m\Nb^\n[G_{\n\a}]  \nonumber\\
                                             && + B^{\a\b}(\Nb_\b K_{\a\m} -\Nb_\m K_{\a\b})
                                                 -B_{\a\m}\Nb_\b K^{\a\b}-K^{\a\b}\Nb_\b B_{\a\m}.
\end{eqnarray}

\item Terms with $\at$:

Regarding $\at$ we have
\[
G^{(2)a_3}_{\alpha\beta}\defi -4R_{\a\m}R^\m_\b + 2 R_{\a \r \m \n} R_\b^{\; \r \m \n} + 4 R_{\a\m\b\n} R^{\m \n}
- 2\N_\beta\N_\alpha R + 4 \Box R_{\a\b} - \frac{1}{2} g_{\a\b}R_{\r\g\m\n}R^{\r\g\m\n}.
\]
All terms have already appeared except for the last one, for which we use (\ref{eq:[RieRie]}).
Straightforward calculations lead to
\begin{eqnarray}
n^\a n^\b [G^{(2)a_3}_{\alpha\beta}] &=&  4 {\cal R}_{\a\b} K^{\a\b} + 4 \Nb_\a \Nb_\b [G^{\a\b}] +4[G_{\a\r}]K^{\a\b}K^\r_\b + 2\Boxb [R]\nonumber \\
&& +4 B^{\a\b}(\RS_{\a\b}-\Rb_{\a\b}+K_{\a\b}K^\r_\r-K_{\a\r}K^\r_\b) , \\
n^\a h^\b_\m [G^{(2)a_3}_{\alpha\beta}] &=& +4 K^\a_\m \Nb^\b[G_{\b\a}] - 4\Nb_\a{\cal R}^\a_\m + 4K^\a_\m \Nb_\a  [R]
                                           - 4 \Nb_\b (B_{\a\m}K^{\a\b}) \nonumber \\
&& +2[R]\Nb_\a K^\a_\m -4B_{\b\m}\Nb_\a K^{\a\b}+4 B^{\a\b}(\Nb_\b K_{\a\m} -\Nb_\m K_{\a\b}).
\end{eqnarray}
\end{itemize} 

Collecting all the above, we finally obtain
\begin{eqnarray}
  &&n^\a n^\b [G^{(2)}_{\alpha\beta}]=\kappa_1\left\{
     bK^\r_\r+\Boxb [R]+\frac{1}{2} \left(\RS-\Rb+(K^\r_\r)^2-K_{\r\sigma} K^{\r\sigma}\right)\right\}\nonumber \\
  &&\qquad+\kappa_2\left\{2 {\cal R}_{\a\b} K^{\a\b} +2\Nb_\a \Nb_\b [G^{\a\b}] +2B^{\a\b}(\RS_{\a\b}-\Rb_{\a\b}+K_{\a\b}K^\r_\r-K_{\a\r}K^\r_\b)\right.\nonumber \\
  &&\qquad \left. +2[G_{\a\m}]K^{\a\b}K_\b^\m+\Boxb [R]\right\} \label{eq:nn[G2]}
\end{eqnarray}
\begin{eqnarray}
  &&n^\a h^\b_\m [G^{(2)}_{\alpha\beta}]=\kappa_1\left\{[R](\Nb_\a K^\a_\m-\Nb_\m K^\r_\r)-\Nb_\m b+K_\m^\a\Nb_\a [R]\right\}\nonumber \\
  &&\qquad+\kappa_2\left\{-2\Nb_\a{\cal R}^\a_\m+2K^\a_\m \Nb^\b[G_{\b\a}] +2 B^{\a\b}(\Nb_\b K_{\a\m}-\Nb_\m K_{\a\b})+2K^\a_\m \Nb_\a [R]\right.\nonumber\\
  &&\qquad+\left.[R]\Nb_\a K^\a_\m-2B_{\a\m}\Nb_\b K^{\a\b}-2K^{\a\b}\Nb_\b B_{\a\m}\right\}.\label{eq:ne[G2]}
\end{eqnarray}

{\bf Remark:} As a final remark, we would like to stress that all the discontinuities computed in this section \ref{sec:nG2} are purely geometrical, and therefore valid in any theory based on a Lorentzian manifold whenever (\ref{Kdisc=0}) holds.

\section{Field equations on the layer $\S$}
\label{sec:field_eqs_S}
Relations (\ref{eq:nn[G2]}) and (\ref{eq:ne[G2]}) are the equations we were looking for, but we wish to rewrite them in terms of the (derivatives of) the energy-momentum quantities supported on $\S$ given in (\ref{tauexc}-\ref{taue}) and (\ref{strength}). Observe, first of all, that the three relations (\ref{rtangent}), (\ref{eq:dRnnt}) and (\ref{eq:dRnnt1}) allow us to rewrite the energy-momentum contents supported on $\S$ (\ref{tauexc}-\ref{taue}) as follows
\bea
\kappa \tau_{\a\b} =-(\kappa_1 + \kappa_2) [R] K_{\a\b}+\kappa_1 b h_{\a\b} + 2 \kappa_2 {\cal R}_{\a\b} ,\label{tauexc1}
\\
\kappa \tau_\a =-(\kappa_1 + \kappa_2)\overline{\N}_\a [R] - 2\kappa_2 \Nb^\r[G_{\r\a}] , \label{tauex1}
\\
\kappa \tau = (\kappa_1 + \kappa_2) [R] K^\r_\r  + 2\kappa_2 K^{\r\sigma} [G_{\r\sigma}], \label{taue1} 
\eea
and using the definition of the double-layer strength (\ref{strength})  the last two here can be rewritten as
\bea
\tau_\a =-\Nb^\r\mu_{\r\a} , \label{tauex2}
\\
\tau =  K^{\r\sigma} \mu_{\r\sigma}. \label{taue2} 
\eea

Now, a direct computation provides the following expressions for some combinations of derivatives of these objects:
\begin{eqnarray}
\kappa \left( \Nb^\b \tau_{\a\b} + K^\r_\r \tau_\a +\Nb_\a \tau \right)  = -(\kappa_1 + \kappa_2) \left(K_\a{}^\b\Nb_\b [R] + [R] (\Nb^\b K_{\a\b}-\Nb_\a K^\r_\r)\right)\nonumber\\
+\kappa_1 \Nb_\a b +2\kappa_2(\Nb^\b {\cal R}_{\a\b}+\Nb_\a(K^{\r\sigma}[G_{\r\sigma}])+K^\r_\r \Nb^\m[G_{\m\a}]),
\eea
\bea
\kappa \left( \tau_{\a\b}K^{\a\b} - \Nb^\a \tau_\a \right) = (\kappa_1 + \kappa_2)(\Boxb [R] - [R] K_{\r\sigma}K^{\r\sigma}) 
\\
+ \kappa_1 b K^\r_\r + 2\kappa_2 (K^{\r\sigma} {\cal R}_{\r\sigma} + \Nb^\a \Nb^\b [G_{\a\b}]).
\end{eqnarray}
Using these, equations (\ref{eq:ne[G2]}) and (\ref{eq:nn[G2]}) become respectively (after some rewriting using (\ref{gauss2}) and (\ref{gauss3}) and (\ref{fe}))
\bean
\kappa\left( n^\alpha h^\r_\b [T_{\a\r}]+ \Nb^\a \tau_{\a\b} + K^\r_\r \tau_\b +\Nb_\b \tau \right)
&=& 2\kappa_2 \left \lbrace K^{\a\r}\overline{\N}_\b [G_{\a\r}] - K^\r_\r  \overline{\N}^\a [G_{\a\b}]\right.\nonumber\\
&& \left. + \overline{\N}_\r ([G^{\a\r}]K_{\a\b}) - \overline{\N}_\r([G_{\a\b}]K^{\a\r})  \right \rbrace, \\
\kappa\left( n^\a n^\b [T_{\a\b}]+ \Nb^\a \tau_\a - \tau_{\a\b}K^{\a\b} \right)&=&  
(\kappa_1 +\kappa_2) [R]\left( n^\a n^\b \RS_{\a\b} + K_{\a\b}K^{\a\b}\right)\nonumber\\&&+2\kappa_2 
[G^{\m\n}]\left (n^\a n^\g \RS_{\a \m\g\n} + K_\m^\r K_{\n\r} \right ).
\eean
Using now the definition of the strength (\ref{strength}) these become
\bean
n^\alpha h^\r_\b [T_{\a\r}]+ \Nb^\a \tau_{\a\b} + K^\r_\r \tau_\b +\Nb_\b \tau 
&=& K^{\a\r}\overline{\N}_\b\mu_{\a\r} - K^\r_\r  \overline{\N}^\a \m_{\a\b} \nonumber \\
&& + \overline{\N}_\r (\m^{\a\r}K_{\a\b}) - \overline{\N}_\r(\m_{\a\b}K^{\a\r})  \\
n^\a n^\b [T_{\a\b}]+ \Nb^\a \tau_\a - \tau_{\a\b}K^{\a\b} &= &
\mu^{\m\n}\left (n^\a n^\g \RS_{\a \m\g\n} + K_\m^\r K_{\n\r} \right).
\eean
Recalling here the relations (\ref{tauex2}) and (\ref{taue2}) between $\tau_\a$ and $\tau$ with the double-layer strength $\mu_{\a\b}$, we finally obtain the following field equations
\begin{empheq}[box=\fbox]{align}
n^\alpha h^\r_\b [T_{\a\r}]+ \Nb^\a \tau_{\a\b} 
&= -\mu_{\a\r} \overline{\N}_\b K^{\a\r} + \overline{\N}_\r (\m^{\a\r}K_{\a\b}) - \overline{\N}_\r(\m_{\a\b}K^{\a\r}) ,
 \label{eq:1} \\
n^\a n^\b [T_{\a\b}] - \tau_{\a\b}K^{\a\b} &=  
\Nb^\a\Nb^\b \m_{\a\b} + \mu^{\m\n}\left (n^\a n^\g \RS_{\a \m\g\n} + K_\m^\r K_{\n\r} \right ).\label{eq:2}
\end{empheq}
These are the fundamental field equations satisfied by the energy-momentum quantities (\ref{tauexc}) and (\ref{strength}) within $\S$. They generalize the classical Israel equations of GR \cite{I} and they are very satisfactory from a physical point of view. They possess an obvious structure with a clear interpretation as energy-momentum conservation relations. There are three type of terms in these relations. The first type is given by the corresponding first summands on the lefthand side. They simply describe the jump of the normal components of the energy-momentum tensor across $\S$. Therefore, they are somehow the main source for the energy-momentum contents in $\S$. The second type of terms are those on the lefthand side involving $\tau_{\a\b}$, the energy-momentum tensor in the shell/layer $\S$. We want to remark that the first equation (\ref{eq:1}) provides the divergence of $\tau_{\a\b}$. Finally, the third type of terms are those on the righthand side, involving the 
 strength
  $\m_{\a\b}$ of a double layer. These terms act also as sources of the energy-momentum contents within $\S$, combined with extrinsic geometric properties of $\S$ and curvature components in the space-time. 

An alternative version of (\ref{eq:1}), after use of the Codazzi equation (\ref{codazzi}), reads
\begin{empheq}[box=\fbox]{align}
n^\alpha h^\r_\b [T_{\a\r}]+ \Nb^\a \tau_{\a\b} 
= \mu^{\a\r} n^\sigma \RS_{\sigma \a \l \r}h^\l_\b +K_{\a\b} \overline{\N}_\r \m^{\a\r} - \overline{\N}_\r(\m_{\a\b}K^{\a\r}) .
\label{eq:2bis}
\end{empheq}
Note that the allowed jumps in the Riemann tensor (\ref{eq:[Rie]}) lead to $ n^\sigma [R_{\sigma \a\l\r}] h^{\a}_\gamma h^\l_\b h^{\r}_\xi = 0$ and therefore the term $\mu^{\a\r} n^\sigma \RS_{\sigma \a\l\r}h^\l_\b$ in the last formula can be written simply as $\mu^{\a\r} n^\sigma R_{\sigma \a\l\r} h^\l_\b$.

\section{Energy-momentum conservation}\label{section:divergence}
The divergence of the lefthand side of the field equations (\ref{fe}) vanishes identically due to the Ricci and Bianchi identities, and therefore, the conservation equation for the energy-momentum tensor $\N_\m T^{\m\n}=0$ follows. In our situation, however, we are dealing with tensor distributions, and with (\ref{fe}) considered in a distributional sense.  The question arises if whether or not the energy-momentum tensor distribution (\ref{emt}) is covariantly conserved. We know that the Bianchi and Ricci identities hold for distributions (see Appendices), 
hence it is expected that the divergence of the $\underline{T}_{\m\n}$ vanishes when distributions are considered. In this section we prove that this is the case, when taking into account the fundamental field equations (\ref{eq:1}) and (\ref{eq:2}). The following calculation can be alternatively seen, therefore, as an independent derivation of (\ref{eq:1}) and (\ref{eq:2}) ---from the covariant conservation of $\underline{T}_{\m\n}$.

From (\ref{emt0}) and (\ref{nablaT1}) we directly get
\be
\nabla^\a \underline{T}_{\a\b} =n^\a [T_{\a\b}] \deltaS +\N^\a (\widetilde T_{\a\b}\deltaS ) + \N^\a \underline{t}_{\a\b}.\label{step}
\ee
Let us first compute the middle term on the righthand side. From the orthogonal decomposition (\ref{emtortog})
$$
\N^\a (\widetilde T_{\a\b}\deltaS ) = \N^\a \left( \left\{\tau_\b+\tau n_\b \right\} n_\a \deltaS\right) +\N^\a \left(\left\{\tau_{\a\b}+\tau_\a n_\b \right\} \deltaS\right)
$$
and using the general formula (\ref{nablaTdelta}) the second summand can be expanded to get 
$$
\N^\a (\widetilde T_{\a\b}\deltaS ) = \N^\a \left( \left\{\tau_\b+\tau n_\b \right\} n_\a \deltaS\right) +\left(\Nb^\a \tau_{\a\b} 
-\tau_{\a\r}K^{\a\r} n_\b +\tau^\a K_{\a\b} +n_\b \Nb^\a \tau_\a \right)\deltaS 
$$
so that with the help of (\ref{tauex2}) we get
\bea
\N^\a (\widetilde T_{\a\b}\deltaS ) &=& \N^\a \left( \left\{\tau_\b+\tau n_\b \right\} n_\a \deltaS\right) \nonumber \\
&+& \left(\Nb^\a \tau_{\a\b} -\tau_{\a\r}K^{\a\r} n_\b- K_{\a\b} \Nb^\r \m_{\r\a}- n_\b  \Nb^\a \Nb^\r\m_{\a\r} \right)\deltaS .\label{step1} 
\eea
With respect to the last term in (\ref{step}), on using definitions (\ref{t}) and (\ref{strength}) we can write for any test vector field $Y^\b$ and using the Ricci identity
\bean
\left<\N^\a \underline{t}_{\a\b} , Y^\b \right> =-\left<\underline{t}_{\a\b} , \N^\a Y^\b \right>=\int_\S \mu_{\a\b} n^\r \N_\r \N^\a Y^\b d\sigma \\
=\int_\S  \left(\m_{\a\b} n^\r \left\{\N^\a \N_\r Y^\b +R^\b{}_{\sigma\r}{}^\a Y^\sigma \right\} \right) d\sigma \\
=\int_\S  \m_{\a\b} n^\r \N^\a \N_\r Y^\b d\sigma - \left<n^\r\m^{\a\sigma} \RS_{\r\a\b\sigma}\deltaS,Y^\b\right>.
\eean
The first integral here can be expanded as
\bean
\int_\S  \m_{\a\b} n^\r \N^\a \N_\r Y^\b d\sigma =\int_\S  \m_{\a\b} \left\{\N^\a (n^\r \N_\r Y^\b) - K^{\a\r} \N_\r Y^\b\right\} d\sigma \\
=\int_\S n^\r\N_\r Y^\b \left(\m_{\a\sigma} K^{\a\sigma} n_\b- \Nb^\a \m_{\a\b} \right) d\sigma
-\int_\S \m_{\a\b} K^{\a\r} \left(\Nb_\r \overline{Y}^\b +(n_\sigma Y^\sigma) K_\r{}^\b \right) d\sigma \\
= \int_\S (\tau n_\b +\tau_\b) n^\r\N_\r Y^\b d\sigma +\int_\S Y^\b \left(\Nb_\r(\m_{\a\b} K^{\a\r}) -n_\b \m_{\a\sigma} K^{\a\r}K_\r{}^\sigma \right)d\sigma \\
=-  \left<\N^\a \left( \left\{\tau_\b+\tau n_\b \right\} n_\a \deltaS\right), Y^\b \right>
+\left<\left(\Nb_\r(\m_{\a\b} K^{\a\r}) -n_\b \m_{\a\sigma} K^{\a\r}K_\r{}^\sigma \right)\deltaS , Y^\b \right>
\eean
so that we arrive at
\be
\N^\a \underline{t}_{\a\b}= -\N^\a \left( \left\{\tau_\b+\tau n_\b \right\} n_\a \deltaS\right)+
 \left(\Nb_\r(\m_{\a\b} K^{\a\r}) -n_\b \m_{\a\sigma} K^{\a\r}K_\r{}^\sigma -n^\r\m^{\a\sigma} \RS_{\r\a\b\sigma}\right)\deltaS . \label{step2}
\ee
Adding up (\ref{step1}) and (\ref{step2}) to (\ref{step}) we finally obtain
\bean
\nabla^\a \underline{T}_{\a\b} &=&\left\{ n^\a [T_{\a\b}] +\Nb^\a \tau_{\a\b} -\tau_{\a\r}K^{\a\r} n_\b + 
\Nb_\r(\m_{\a\b} K^{\a\r}) \right. \\
&&  \left. - n_\b \m_{\a\sigma} K^{\a\r}K_\r{}^\sigma -n^\r\m^{\a\sigma} \RS_{\r\a\b\sigma} - K_{\a\b} \Nb^\r \m_{\r\a}-  n_\b  \Nb^\a \Nb^\r\m_{\a\r}  \right\} \deltaS .
\eean
The fundamental equations (\ref{eq:2}) and (\ref{eq:2bis}) prove the vanishing of this expression leading to
$$
\nabla^\a \underline{T}_{\a\b} =0
$$
as claimed. As remarked in \cite{Senovilla14,Senovilla15}, this calculation shows that the double-layer energy-momentum distribution $\underline{t}_{\a\b}$ is essential to keep energy-momentum conservation. Without the double-layer contribution the total energy-momentum tensor distribution $\underline{T}_{\a\b}$ would not be covariantly conserved. 

\section{Matching hypersurfaces, thin shells and double layers}
\label{sec:8}
Once we have discussed the junction in the case of gravity theories with quadratic terms, and have obtained the corresponding field equations on $\S$, we are in disposition to analyze their consequences. Before entering into this discussion, it is convenient to remark the following important result.
\begin{result}\label{NoDL}
If there is no double layer (that is $\m_{\a\b}=0$), then there can be neither external flux momentum $\tau_\a$ nor external pressure/tension $\tau$.
\end{result}
This follows directly from expressions (\ref{tauex2}) and (\ref{taue2}). In other words, there exist non-vanishing external flux momentum and/or external pressure/tension {\em only if} there is a double layer. 

Thus, there are three levels of junction depending on whether or not thin shells and/or double layers are allowed. We will term them as:
\begin{itemize}
\item {\em Proper matching}: this is the case where the matching hypersurface $\S$ does not support any distributional matter content, describing simply an interface with jumps in the energy-momentum tensor, so that there are neither thin shells nor double layers. This situation models, for instance, the gravitational field of stars (non-empty interior) with a vacuum exterior. Or the case of vacuoles in cosmological surroundings.
\item {\em Thin shells, but no double layer}: This is an idealized situation where an enormous quantity of matter is concentrated on a very thin region mathematically described by $\S$ but no double layer is permitted to exist. Thus, delta-type terms proportional to $\deltaS$ are allowed, and the expression (\ref{tauexc}) provides the energy-momentum tensor of the thin shell. However, from Result \ref{NoDL} the other possible quantities (\ref{tauex}) and (\ref{taue}) accompanying $\deltaS$ vanish identically. This situation is analogous to that  in GR where only (\ref{tauexc}) appears. The main difference with a generic quadratic gravity arises in the explicit expression for (\ref{tauexc}), as the field equations turn out to adopt the same form. 
\item {\em Double layers}: this is the general case with no further assumptions, which describes a large concentration of matter on $\S$, as in the previous case, but accompanied with a brusque jump in the curvature of the spacetime. Still, there are several sub-possibilities depending on the vanishing or not of any of  (\ref{tauexc}), (\ref{tauex}) or (\ref{taue}). There is also an extreme possibility, that we term a {\em pure double layer}, where the thin shell is not present but the double layer is: this is the case with all (\ref{tauexc}), (\ref{tauex}) and (\ref{taue}) vanishing but a non-vanishing (\ref{t}). Nothing like any of these different possibilities can be described in GR. 
\end{itemize}
We classify the junction condition for these different cases in turn.

\subsection{Thin shells without double layer}
\label{subsecion:thinshells}
From (\ref{t}) follows that the strength of the double layer $\m_{\a\b}$ must be set to zero, and thus from (\ref{strength}) we have
\be
(\kappa_1 +\kappa_2) [R] h_{\a\b} +2\kappa_2 [G_{\a\b}]=0 \hspace{1cm} \Longrightarrow \hspace{3mm} (\kappa_2 +n\kappa_1) [R]=0, \label{mu=0}
\ee
which implies that $\tau$ and $\tau_\alpha$ both vanish (see Result \ref{NoDL}).
 Hence, only the tangential part of the distributional energy momentum tensor on $\Sigma$ survives, given explicitly by (\ref{tauexc1}). Its trace, upon using (\ref{traceRcal}), reads
\be
\kappa \tau_\alpha^\alpha = (n\kappa_1 + \kappa_2)b - K^{\alpha \beta}\mu_{\alpha \beta} = (n\kappa_1 + \kappa_2)b. \label{trace_emsigma}
\ee 
The equations (\ref{eq:1}) and (\ref{eq:2}) in this case read
\be
n^\a h^\r_\b [T_{\a\r}]= -\Nb^\a \tau_{\a\b},\qquad
n^\a n^\b [T_{\a\b}]= \tau_{\a\b} K^{\a\b}. \label{Israelcond_thinshell}
\ee
Observe that, remarkably, these are identical with the Israel conditions derived in GR.

We have to distinguish whether $\kappa_2=0$ or not. 
\begin{itemize}
\item $\kappa_2 \neq 0$.

If $(n\kappa_1 + \kappa_2) \neq 0$ relations (\ref{mu=0}) imply that $[R]=0$ and $[G_{\a\b}] = 0$ in full.
Direct consequences are $[R_{\a\b}] = [R_{\a\b\m\n}] = 0$, and the discontinuities in the derivatives are given by
\begin{equation}
\left[\nabla_\mu R_{\a\b\lambda \nu}\right] =(n_\a n_\lambda {\cal{R}}_{\beta \nu} -n_\a n_\nu \mathcal{R}_{\beta \lambda} -n_\beta n_\lambda \mathcal{R}_{\a \nu} +n_\beta n_\nu \mathcal{R}_{\a \lambda})n_\mu, \label{noshellsDRiemann}
\end{equation}
for some symmetric tensor $\mathcal{R}_{\a\b}$ tangent to $\Sigma$.
From (\ref{para_sacar_b}) we get $b = 2 \mathcal{R}_\rho^\rho$ and therefore the energy-momentum tensor (\ref{tauexc}) on $\Sigma$ just reads 
\be
\kappa \tau_{\a\b} =  \kappa_1 b h_{\a\b} + 2 \kappa_ 2 \mathcal{R}_{\a\b}. \nonumber
\ee

With regard to the exceptional possibility $n \kappa_1 + \kappa_2 = 0$, equation (\ref{mu=0}) implies in particular that the tensor $B_{\a\b}$ is proportional to the first fundamental form. The explicit relation reads
\begin{equation}
B_{\a\b} = \frac{1}{2n}[R] h_{\a\b}, \nonumber
\end{equation}
which for the discontinuity of the Riemann tensor produces
\begin{equation}
\left[R_{\a\b\lambda \mu }\right] = \frac{[R]}{2n}\left(n_\a n_\lambda h_{\b \mu}-n_\lambda n_\beta h_{\a \mu}-n_\mu n_\a h_{\b \lambda}+n_\mu n_\beta h_{\a \lambda}\right). \label{noshellparticularRie}
\end{equation}
Taking contractions in this last expression we find the allowed  jumps in the Ricci and Einstein tensor 
\begin{equation}
\left[R_{\a\b}\right] =\frac{[R] }{2} \left(\frac{1}{n} h_{\a\b}+ n_\a n_\b \right)  \Rightarrow  \left[G_{\a\b}\right]=\frac{1-n}{2n}[R] h_{\a\b}. \label{noshellparticularGandRic}
\end{equation}
Note $[R]$ is the only degree of freedom allowed for the discontinuities of the curvature tensors.

The remaining allowed discontinuities of the derivative of the Ricci tensor are encoded in $r_{\a\b} = n^\mu [\nabla_\mu R_{\a\b}]$, so that
\begin{equation}
\left[\nabla_\mu R_{\a\b}\right] = r_{\a\b} n_\mu +\frac{1}{2}\left(n_\a n_\b  + \frac{1}{n}h_{\a\b}\right)\overline{\nabla}_\mu [R] + \left(\frac{1-n}{2n} \right) [R] \left(n_\a K_{\b \mu} + n_\b K_{\a\mu}\right). \label{noshellparticularDRicci}
\end{equation}
Recalling that $b=r^\a_\a = n^\r [\nabla_\r R]$ the explicit form of the energy momentum tensor on $\Sigma$ can be obtained from (\ref{tauexc1}). Due to (\ref{trace_emsigma}), $\tau_{\a\b}$ is traceless. Nevertheless, the relevance of this exceptional case is probably marginal, as the coupling constants satisfy a dimensionally dependent condition.

\item $\kappa_2 = 0$.

We have to assume then that $\kappa_1 \neq 0$, as otherwise all the terms  (\ref{tauexc}), (\ref{tauex}) and (\ref{taue}) vanish identically and thus there are no thin shells. 
Let us also recall that $\ad$ and $\at$ are assumed not to vanish simultaneously, as that case was fully analysed in \cite{Senovilla13,Senovilla14,Senovilla15}, so it would be more precise to label this case as $\ad = -4\at$ with $\au \neq \at$.

This case reduces to the condition $[R]=0$ (see (\ref{mu=0})). All the remaining jumps on the curvature tensor and its derivatives are allowed.
The energy-momentum tensor on $\S$ is just given by
\begin{equation}
\kappa \tau_{\a\b}=\kappa_1 b h_{\a\b}, \label{noshellk2zerotau}
\end{equation}
with $b=n^\a[\nabla_\a R]$,
and therefore the thin shell $\S$ only contains, at most,
a ``cosmological constant''-type of matter content.

\end{itemize}

\subsection{Proper matching hypersurface}
In addition to the requirement imposed in the previous case of thin shells, we demand now that the full $\tilde T_{\a\b}$ vanishes. Thus we have to add $\tau_{\a\b} = 0$ to the conditions discussed in the previous Subsection \ref{subsecion:thinshells}.  
In general, from (\ref{Israelcond_thinshell}) we have
\be
n^\a [T_{\a\b}]=0 \label{Israelcond}
\ee
which adopt exactly the same form as in GR and we call the generalized Israel conditions. They imply the continuity of the normal components of the energy-momentum tensor across $\S$.

Again, we have to distinguish two cases depending on whether $\kappa_2$ vanishes or not.
\begin{itemize}
\item $\kappa_2\neq 0$.

If $(n\kappa_1 + \kappa_2) \neq 0$, we already know from the previous section that $[R] = 0$ and $[G_{\a\b}] = 0$. The trace relation (\ref{trace_emsigma}) provides $b=0$ and moreover $\tau_{\a\b}=0$ implies, via (\ref{tauexc1}), ${\cal R}_{\a\b}=0$. Plugging this information into (\ref{noshellsDRiemann}) it follows that the derivatives of the curvature tensors do not present discontinuities. 
\begin{result}
In the generic case with $4\at +\ad\neq0$ and $4\at +(1+n)\ad +4n\au \neq 0$, the full set of matching conditions amount to those of GR (agreement of the first and second fundamental forms on $\S$) plus the agreement of the Ricci tensor and its first derivative on $\S$:
\be
[R_{\a\b}]=0 , \hspace{2cm} [\nabla_\r R_{\a\b}]=0. \label{junction}
\ee
\end{result}
This actually implies that the full Riemann tensor and its first derivatives have no jumps across $\S$:
$$
[R_{\a\b\l\m}]=0 , \hspace{2cm} [\nabla_\r R_{\a\b\l\m}]=0.
$$

With regard to the exceptional possibility $\kappa_2 +n \kappa_1 =0$, the curvature tensors satisfy (\ref{noshellparticularRie}) and (\ref{noshellparticularGandRic}). Now $\tau_{\a\b}=0$ provides
$$
 {\cal R}_{\a\b} =\frac{1}{2n} \left( (n-1) [R] K_{\a\b} +b h_{\a\b}\right) ,
$$
and thus $r_{\a\b}=n^\r [\nabla_\r R_{\a\b}]$ gets determined in terms of $[R]$ and $b$, so that (\ref{noshellparticularDRicci}) for  $[\N_\m R_{\a\b}]$  reads  
\begin{eqnarray}
\left[\nabla_\mu R_{\a\b}\right] &=& \left(\frac{1}{2n} \left( (n-1) [R] K_{\a\b} +b h_{\a\b}\right) -2n_{(\a} \overline{\nabla}_{\b)} [R]  +\left(\frac{b}{2} + \frac{1-n}{2n}[R]K_\r^\r \right)n_\a n_\b \right)n_\mu \nonumber \\
&+&\frac{1}{2}\left(n_\a n_\b  + \frac{1}{n}h_{\a\b}\right)\overline{\nabla}_\mu [R] + \left(\frac{1-n}{2n} \right) [R] \left(n_\a K_{\b \mu} + n_\b K_{\a\mu}\right). \nonumber
\end{eqnarray}

Hence, the entire set of discontinuities of the Riemann tensor and its first derivative can be written just in terms of $[R]$ and $b=n^\r [\N_\r R]$, which remain as two free degrees of freedom. As mentioned before, this case is probably irrelevant due to its defining condition depending on the dimension $n$.

\item $\kappa_2 =0$ but $\kappa_1\neq 0$. 

From the results from the previous section we know that $[R]=0$ and the energy momentum on $\Sigma$ is given by (\ref{noshellk2zerotau}). Thus, for a proper matching we find $b=0$. 
The discontinuity in the derivative is
\bean
[\N_\m R_{\a\b}] &=& n_\m\left([R_{\r\n}]K^{\r\n} n_\a n_\b -2 \overline{\N}^\r [R_{\r(\b}]n_{\a)} + {\cal R}_{\a\b}\right) \\
&+& \overline{\N}_\m [R_{\a\b}] -2K_\m^\r [R_{\r(\a}] n_{\b)} ,
\eean
where also ${\cal R}^\r_\r = -K^{\a\b}[R_{\a\b}]$. 

\item $\kappa_1 = \kappa_2 = 0$.

Or equivalently $\au = \at = -\ad/4$. In this case all the terms  (\ref{tauexc}), (\ref{tauex}) and (\ref{taue}) and (\ref{t}) vanish identically and thus there are no further restrictions other than $[K_{ab}] = 0$. The junction conditions are just the same as in GR. This is the case where the quadratic part of the Lagrangian (\ref{lag}) is the Gauss-Bonnet term \cite{Lovelock71}.

\end{itemize}

\subsection{The double layer fauna; pure double layers}
The generic occurrence in quadratic gravity, as shown above, is that any thin shell comes accompanied by a double layer, which in turn generically implies the existence of non-zero external pressure/tension and external flux momentum. However, there are several special possibilities in which one of these quantities, or all, disappear. This gives rise to a fauna of different kinds of double layers. There is also the possibility that the double layer term (\ref{t}) is non-zero while the remaining distributional part in the energy-momentum tensor, that is $\tilde T_{\a\b}\deltaS$, vanishes. In other words, a double layer without a classical thin shell. We call such a case a {\em pure double layer}. In the rest of this section we explore this novel possibility.

For pure double layers, the vanishing of the external pressure $\tau$ plus the energy flux $\tau_\a$ first imply,
by virtue of (\ref{tauex}) and (\ref{tauex2})
\be
\mu_{\a\b}K^{\a\b}=0,\quad \Nb^\r \mu_{\r\a}=0.
\label{pure_DB_1}
\ee
In particular, then, the double layer strength is conserved. 

The first equation in (\ref{pure_DB_1}) yields
\begin{equation}
(\kappa_1 +\kappa_2)[R] K^\sigma_\sigma + 2\kappa_2 K^{\rho \sigma}[G_{\rho \sigma}] = 0\label{dlnoshell2}
\end{equation}
while the second gives
\be 
(\kappa_1 +\kappa_2) \overline{\N}_\a [R] + 2\kappa_2 \overline{\nabla}^\rho [G_{\rho\a}]= 0. \label{dlnoshell3}
\ee
Equation (\ref{dlnoshell2}) combined with the vanishing of the trace of $\tau_{\alpha \beta}$ provides
\begin{equation}
(\kappa_1 n + \kappa_2)b=0 \label{dlnoshell1}
\end{equation}
so that, generically --- $n \kappa_1 + \kappa_2 \neq 0$ --- one has $b=0$. A first consequence is that the jump in the derivative of the Ricci scalar is now tangent to $\Sigma$ and fully determined by the tangent derivative of $[R]$
\be 
[\nabla_\a R] = \overline{\nabla}_\a [R].\label{nablaR=nablaR}
\ee 
The vanishing of $\tau_{\alpha \beta}$, using (\ref{tauexc}), is now equivalent to
\begin{equation}
 \kappa_2 {\cal R}_{\a\b} = (\kappa_1 + \kappa_2)\frac{[R]}{2} K_{\alpha \beta}. \label{dlnoshell4}
\end{equation}
The expression for the discontinuity of the normal derivative of the Ricci tensor  has to be studied depending on $\kappa_2$ vanishing or not.
\begin{itemize}
\item $\kappa_2\neq 0$

The relations above allow us to write the discontinuity of the normal derivative of the Ricci tensor as
\be 
r_{\a\b} = \frac{1}{2}\left(1+ \frac{\kappa_1}{\kappa_2} \right)([R]K_{\a\b} + n_\b \overline{\nabla}_\a [R] + n_\a \overline{\nabla}_\b [R] - K[R] n_\a n_\b), \nonumber
\ee

whereas the tangent part of the derivative keeps its original form given in (\ref{eq:[d_diff_ricci]}).

\item $\kappa_2=0$ (and $\kappa_1 \neq 0$).

Equations (\ref{dlnoshell3}) and (\ref{dlnoshell4}) read
\be 
\overline{\nabla}_\a [R] = 0, \quad [R]K_{\a\b}=0, \label{eq:dlk1zero}
\ee
and (\ref{dlnoshell2}) is automatically satisfied. Thus, (\ref{nablaR=nablaR}) implies
 $[\nabla_\a R] = 0$. 
 Observe that since $\kappa_2=0$, (\ref{strength}) establishes that the strength of the double layer is proportional to $[R]$. Hence, in order to have a nonzero $\mu_{\a\b}$, $[R]$ cannot vanish.
 Then $K_{\a\b}=0$ necessarily, and the allowed jumps are encoded in $[R_{\a\b}]$ and $r_{\a\b}$. 
\end{itemize}

For completeness, we provide finally the formulas for the exceptional case $n\kappa_1 + \kappa_2=0$ ---discarding $\kappa_1 = \kappa_2 = 0$ for which the double layer simply disappears. The equations $\tau = 0$, $\tau_\a= 0$ and $\tau_{\a \b} = 0$ result, respectively, in 
\begin{eqnarray}
	(1-n) [R] K^\b_\b - 2n K^{\a\b}[G_{\a\b}] &=& 0, \nonumber \\
	(1-n)\overline{\nabla}_\a [R] -2n \overline{\nabla}^\rho [G_{\rho \a}] &=&0, \nonumber \\
	(1-n)[R]  K_{\a\b} - b h_{\a\b} + 2n \cal{R}_{\a \b} &=& 0. \nonumber
\end{eqnarray}
While the third equation provides $\mathcal{R}_{\a \b}$, the first two constitute
constraints on the allowed jumps of the Ricci tensor that should be analysed
in each particular situation.
In all cases, the allowed discontinuity in the derivative of the Ricci tensor can be written as
\begin{eqnarray}
r_{\a\b} &=& - \frac{1}{2n} \left((1-n)[R]K_{\a\b} - bh_{\a\b}\right)  -\frac{1-n}{2n} \left(n_\b \overline{\nabla}_\a [R] + n_\a \overline{\nabla}_\b [R]\right)  \nonumber\\
&& +\frac{1}{2}\left(b + \frac{1-n}{n}[R]K^\r_\r \right)n_\a n_\b . \nonumber 
\end{eqnarray}
Observe that now the strength of the double layer is traceless, $\mu^\r_\r=0$
(see e.g.(\ref{tracestrength})).

\section{Consequences}
\label{sec:consequences}
The proper matching conditions in GR are the agreement of the first and second fundamental forms on $\S$. Therefore, any matching hypersurface in GR satisfies (\ref{Kdisc=0}), and the allowed jumps in the energy-momentum tensor are equivalent to non-vanishing discontinuities of the Ricci (and Riemann) tensor. Thus, in GR properly matched space-times will generally have $[R_{\a\b}]\neq 0$.

This simple known fact implies that any GR-solution containing a proper matching hypersurface {\em will contain a double layer and/or a thin shell at the matching hypersurface if the true theory is quadratic} ! 

At least two relevant consequences follow from this fact: (i) generically, matched solutions in GR are no longer solutions in quadratic theories; and (ii) if any quantum regimes require, excite or switch on quadratic terms in the Lagrangian density, then GR solutions modelling two regions with different matter contents will develop thin shells and double layers on their interfaces. Let us elaborate.

Consider, for instance, the case of a perfect fluid matched to a vacuum in GR. As is well known, the GR matching hypersurface is determined by the condition that
$$
p^{GR} |_\S =0
$$
where $p^{GR}$ is the isotropic pressure of the fluid in GR. It follows that the Ricci tensor has a discontinuity of the following type
$$
[G_{\a\b}] = \left. \kappa \varrho^{GR} u_\a u_\b \right|_\S, \hspace{1cm} [R_{\a\b}] =\left. \kappa \varrho^{GR}  \left(u_\a u_\b +\frac{1}{n-1} g_{\a\b}\right)\right|_\S
$$
$u^\a$ being the unit velocity vector of the perfect fluid. Therefore, using (\ref{tauexc1}-\ref{taue1}) and (\ref{t}) we see that the very same space-time has, in any quadratic theory of gravity, an energy-momentum tensor distribution with all type of thin-shell and double-layer terms.

Imagine the situation of a collapsing perfect fluid (to form a black hole, say) with vacuum exterior. Then one can use any of the known solutions in GR to describe this situation ---the reader may have in mind, for instance, the Oppenheimer-Snyder model. The GR solution describes this process accurately in the initial and intermediate stages, when the curvature of the space-time is moderate and the values of $\au R^2$, $ \ad R_{\a\b}R^{\a\b}$ and $\at R_{\a\b\m\n}R^{\a\b\m\n}$ for instance, or other similar quantities, are small enough to render any quadratic terms in the Lagrangian totally negligible. However,  as the collapse proceeds and one approaches the black hole regions ---and later the classical singularity--, regimes with very high curvatures are reached. Then, the quadratic terms coming from any quantum corrections (be they from string theory counter-terms, or any other) to the Einstein-Hilbert Lagrangian start to be important, and actually to dominate, the curvature being  enormous. In this regime, the original  matching hypersurface becomes actually a thin double layer.

Of course, the description of a global space-time via a matching is an approximation, and also the use of tensor distributions is also just another approximation to a real situation where a gigantic quantity of matter can be concentrated around a very thin region of the space-time. Nevertheless, both approximations are satisfactory in the sense that they are believed to actually mimic a realistic situation where the layer is thick and the jumps in the energy variables are extremely big, but finite. If this is the case, then the above reasoning seems to imply that, {\em if quadratic theories of gravity are correct}, at least in some extreme regimes, then a huge concentration of matter will develop around the interface of the interior and the exterior of the collapsing star. And this huge concentration will generically manifest as a shell with double-layer properties.

\section*{Acknowledgments}
The authors are supported by FONDOS FEDER under grant FIS2014-57956-P (Spanish government), and by projects GIU12/15 IT592-13 (Gobierno Vasco) and UFI 11/55 (UPV/EHU). JMMS is also supported by project P09-FQM-4496 (J. Andaluc\'{\i}a---FEDER). BR acknowledges financial support from the Basque Government grant BFI-2011-250.

\section*{Appendices}
\appendix
\section{Tensor distributions}\label{App:A}
Let ${\cal D}(V)$ be the set of sufficiently differentiable tensor fields of any order with {\em compact
support} on the (oriented) Lorentzian manifold $(V,g)$ and denote by ${\cal D}^q_p$ the subset of $p$-covariant
$q$-contravariant such tensor fields. The elements of these sets are called {\em test tensor fields}. The definition of tensor distributions is as follows (see, for instance \cite{L,MS,T})
\begin{definition}[Tensor distribution]\label{distributions}
A $p$-covariant $q$-contravariant  tensor distribution $\chi^q_p$ is a linear and
continuous functional
\begin{eqnarray*}
\chi^q_p : {\cal D}^p_q & \rightarrow & \mathbb{R} \\
Y^p_q & \rightarrow & \chi^q_p \left ( Y^p_q \right ) \defi \left < \chi^q_p,
Y^p_q \right > \enspace .
\end{eqnarray*}
\end{definition}
The set of tensor distributions of the same type constitutes
a $\mathbb{R}$-vector space with the natural definitions. This set also contains all locally integrable tensor fields: any
 $p$-covariant $q$-contravariant tensor field $T^q_p$  defines a unique tensor distribution ${\underline{T}}^q_p$ by means of
\begin{eqnarray*}
{\underline{T}}^q_p : {\cal D}_q^p & \rightarrow & \mathbb{R} \\ Y^p_q  &
\rightarrow & \left <{\underline{T}}^q_p, Y^p_q \right >\equiv
\int_V \left ( T^q_p,Y^p_q \right ) \mbox{\boldmath$\eta$}
\end{eqnarray*}
where $\bm{\eta}$ is the volume element $(n+1)$-form and $\left (T,Y \right )$ is a shorthand for the scalar
$$
( T^q_p,Y^p_q ) \defi T_{\mu_1\dots\mu_p}^{\nu_1\dots\nu_q} Y^{\mu_1\dots\mu_p}_{\nu_1\dots\nu_q}
$$
Observe that, following \cite{MS}, we use the
convention of distinguishing between a tensor field and its associated tensor distribution by using an underline. We will also put an underline on some other tensor distributions, not associated to tensor fields, to emphasize their distributional character.
\begin{definition}[Tensor distribution components]\label{components}
The components of a $(p,q)$-tensor distribution $\chi$ relative to dual bases $\{\bm{\theta}^\mu\}$, $\{\vec{e}_\mu\}$ are
scalar distributions $\chi^{\alpha_1\cdots\alpha_q}_{\beta_1\cdots\beta_p}$
defined by
\begin{eqnarray*}
\left < \chi^{\alpha_1\cdots\alpha_q}_{\beta_1\cdots\beta_p}, Y \right >
\defi \left < \chi^q_p, Y\, \bm{\theta}^{\alpha_1}\otimes\cdots\otimes 
 \bm{\theta}^{\alpha_p}
\otimes \vec{e}_{\beta_1}\otimes\cdots\otimes 
\vec{e}_{\beta_q} \right >
\end{eqnarray*}
where $Y$ is any test function. 
\end{definition}
It easily follows
\be
\left < \chi^q_p, Y^p_q \right > = \left < \chi^{\alpha_1\cdots\alpha_q}_{
\beta_1\cdots\beta_p},Y_{\alpha_1\cdots\alpha_q}^{\beta_1\cdots\beta_p}
\right > \enspace .\label{InComp}
\ee
Contraction of indexes are then well-defined and independent of the basis for tensor distributions. 

A tensor distribution $\chi^q_p$ is said to vanish on an open set $U\subset V$ if 
$$
\left < \chi^q_p,Y^p_q \right >=0
$$
for {\em all} test tensor fields with support contained in $U$.
\begin{definition}[Support of tensor distributions]
The support of a tensor distribution $\chi^q_p$ is the complement in $V$ of the union of all open sets where $\chi^q_p$ vanishes.
\end{definition}
The support of a tensor distribution is always a closed set.

\begin{definition}[Tensor product by tensor fields]\label{otimes}
The tensor product of a tensor distribution $\chi_p^q$ by a tensor field $T^r_s$ ---defined on a neighbourhood of the support of $\chi^q_p$--- is the
$(p+s)$-covariant $(q+r)$-contravariant tensor distribution acting as
\begin{eqnarray*}
\left < T^r_s\otimes\chi^q_p, Y^{s+p}_{r+q} \right > \equiv \left < 
\chi^q_p, \left ( T,Y\right )^p_q \right>
\end{eqnarray*}
where $(T,Y)^p_q\in {\cal D}_q^p$ is given, in any basis, by
$$
(T,Y)^{\mu_1\dots\mu_p}_{\nu_1\dots\nu_q} \defi T^{\rho_1\dots\rho_r}_{\sigma_1\dots\sigma_s}\, Y^{\sigma_1\dots\sigma_s\mu_1\dots\mu_p}_{\rho_1\dots\rho_r\nu_1\dots\nu_q}
$$
\end{definition}
Sometimes, for this product to make sense it is enough that the tensor field $T^r_s$ is defined just on the support of $\chi^q_p$, however this will not be the case when derivatives are involved. Care must be taken with this problem.

Tensor distributions can be differentiated (by acting on differentiable elements of ${\cal D}(V)$). The main definition is as follows.
\begin{definition}[Covariant derivative of tensor distributions]\label{derivative}
The covariant derivative $ \nabla \chi^q_p$ of a $(p,q)$-tensor distribution $\chi^q_p$ is the $(p+1,q)$-tensor distribution defined by
\begin{eqnarray*}
\left < \nabla \chi^q_p, Y^{p+1}_q \right > \defi -\left < \chi ^q_p, (DY)^p_q
\right >
\end{eqnarray*}
where $\left (DY\right )^{\alpha_1\cdots\alpha_p}_{\beta_1\cdots\beta_q}\defi
\nabla_{\mu}Y^{\mu\alpha_1\cdots\alpha_p}_{\beta_1\cdots\beta_q}$.
\end{definition}
This definition is well posed in the sense that it recovers the usual covariant derivative for differentiable tensor fields:
$$
\nabla\underline{T}= \underline{\nabla T} .
$$
The components of $\nabla \chi^q_p$ in any basis are denoted by $\nabla_{\mu} \chi^{\alpha_1\cdots\alpha_q}_{\beta_1\cdots\beta_p}$.

\section{Matching hypersurfaces and thin shells}\label{App:B}
Consider the case where the spacetime $(V,g)$ contains a hypersurface $\Sigma$ such that the metric 
tensor (or its first derivatives) may be not differentiable across $\S$. In that case, its derivatives and the
Christoffel symbols may not exist at points of $\Sigma$ ---but they do
outside this hypersurface. Similarly, the curvature tensors and their derivatives will not be defined on $\S$ as tensor fields. However, given that the metric tensor $g$ is continuous, it defines a tensor distribution $\underline g$ which can be differentiated {\em in the distributional sense}. Thus, one can define the curvature and its derivatives as tensor distributions. 

To that end, define the $\Sigma$-step function $\theta : V \rightarrow  \mathbb{R} $ as
\bea
\theta =\left\{
\begin{array}{ccc}
1 &  & V^+\\
1/2 & \mbox{on} & \Sigma \\
0 &  & V^-
\end{array}\right. \label{theta}
\eea
This function is locally integrable so that it defines a scalar distribution $\otheta$:
\begin{eqnarray*}
\left < \otheta, Y \right > = \int_{V^{+}} Y \mbox{\boldmath$\eta$}  
\end{eqnarray*}
whose covariant derivative is a one-form distribution with support on $\S$ acting as
$$
\left<\nabla\otheta ,\vec Y\right>=-\left<\otheta,D\vec Y\right>=-\int_{V^+} \nabla_\mu Y^\mu \bm{\eta}= \int_{\Sigma}
Y^{\mu}d\sigma_{\mu}=\int_{\Sigma} Y^{\mu}n_{\mu} d\sigma
$$
where $d\sigma_{\mu}=n_\mu d\sigma$ is oriented from $V^{-}$ towards $V^{+}$, $n_\mu$ is the unit normal to $\S$, and thus $d\sigma$ is its canonical volume element $n$-form. It arises a natural scalar distribution $\deltaS$ with support on $\S$ defined by
\be
\left < \deltaS, Y \right >\defi \int_{\Sigma} Y d\sigma \enspace .\label{deltaS}
\ee
This distribution can act on every locally integrable test function defined at least at
the points of $\S$. Similarly, it can be multiplied by any locally integrable tensor field defined at least on $\S$. Observe in particular, from the above, that 
\be
\nabla_\mu\,   \underline{\theta} = n_\mu\,  \deltaS .\label{nablatheta}
\ee

Let $T$ be any $(p,q)$-tensor field which (i) may be discontinuous across $\Sigma$, (ii) is differentiable
on $V^{+}$ and $V^{-}$, and (iii) such that $T$ and its covariant derivative have
definite limits on $\Sigma$ coming from both $V^{+}$ and $V^{-}$. We will use the notation
$T^{\pm}$ for the restriction of $T$ to $V^{\pm}$ respectively. The tensor distribution associated to such a $T$, which exists because the tensor is locally integrable, is
\be
\underline{T}=T^{+} \otheta +T^{-} \left( \1-\otheta \right ) \enspace . \label{Tdist}
\ee
By definition, we also take
$$
T\defi T^{+} \theta +T^{-} \left( 1-\theta \right )
$$
as the tensor field defined {\em everywhere}, and thus at each point of $\S$
\be
T^\S\defi T|_\S = \frac{1}{2} \left(\mathop{\lim} \limits_{x \mathop \to \limits_{V^{+}}\S} T^{+}(x) +
\mathop{\lim} \limits_{x \mathop \to \limits_{V^{-}}\S} T^{-}(x) \right)\, .\label{TonS}
\ee
The covariant derivative of $\underline T$ can be shown to be \cite{MS}
\be
\underline{\nabla T} = \nabla T^+ \otheta +\nabla T^- (\1 -\otheta) +\bm{n}\otimes \left[T\right] \deltaS \label{nablaT}
\ee
where $\left[ T\right ]$ is a a $(p,q)$-tensor field defined only on $\Sigma$, that we call the ``jump'' or the ``discontinuity'' of $T$ at $\Sigma$ and is defined as
\be
\forall q \in \Sigma, \hspace{1cm} \ \left[ T \right ] (q) \equiv
 \mathop{\lim} \limits_{x \mathop \to \limits_{V^{+}}q} T^{+}(x) -
\mathop{\lim} \limits_{x \mathop \to \limits_{V^{-}}q} T^{-}(x) \enspace .
\label{discont}
\ee
The index version of (\ref{nablaT}) reads
\be
\nabla_\mu\, \underline{T}^{\a_1\dots\a_q}_{\b_1\dots\b_p} = \nabla_\mu T^{+\a_1\dots\a_q}_{\b_1\dots\b_p} \otheta
+\nabla_\mu T^{-\a_1\dots\a_q}_{\b_1\dots\b_p} (\1 -\otheta )+\left[T^{\a_1\dots\a_q}_{\b_1\dots\b_p} \right] n_\mu \deltaS.
\label{nablaT1}
\ee
Formula (\ref{nablatheta}) is just a particular case of this general formula.

It must be noticed that formula (\ref{nablaT}), or (\ref{nablaT1}), is precisely the formula one would derive by using a naif calculation starting from (\ref{Tdist}), applying Leibniz rule and using (\ref{nablatheta}). However, such a naif approach {\em cannot be used} when the tensor distribution to be differentiated involves non-tensorial distributions ---such as $\deltaS$. For instance, one may be tempted to compute the second covariant derivative of $\otheta$ starting from (\ref{nablatheta}) and write
$$
\nabla_\nu \nabla_\mu \otheta \,\, \xcancel{=} \,\,  \nabla_\nu n_\mu \, \deltaS + n_\mu \nabla_\nu \deltaS \hspace{1cm} \mbox{wrong!!}
$$
Neither term on the righthand side is well defined due the the fact that $n_\mu$ exists only on $\S$ and therefore its derivatives non-tangent to $\S$ are not defined at all. Nevertheless, $\nabla_\nu \nabla_\mu \otheta $ is certainly well defined as a distribution, and one can compute an explicit formula by following strictly the rules of tensor-distribution derivation and multiplication. This is done in full generality in Appendix \ref{App:D3}, formula (\ref{nablaTdelta}),  which provides the sought result in (\ref{nablandelta}) as well as later the formula (\ref{nabladelta}). 

\section{Curvature tensor distributions}\label{App:C}
We are now prepared to compute the connection and the Riemann tensor in the distributional sense. Recall that the  metric $g$ is continuous across $\S$ and can be written as a function or as a distribution
\begin{eqnarray*}
g= g^{+}  \theta+g^{-} \left( 1- \theta \right) , \hspace{1cm} \underline{g}=g^{+}  \otheta + g^{-} \left( \1- \otheta \right) .
\end{eqnarray*}
As in the general rule (\ref{nablaT1}) one deduces
$$
\partial_\a \underline{g}{}_{\mu\nu} =\partial_\a g^+_{\mu\nu} \otheta + \partial_\a g^-_{\mu\nu} (\1-\otheta)
$$
from where, using the standard formula, one can construct the Christoffel symbols as distributions
\begin{eqnarray}
{\underline{\Gamma}}^{\alpha}_{\beta\gamma}= {\Gamma^{+}}^{\alpha}_{\beta\gamma}  \otheta +
{\Gamma^{-}}^{\alpha}_{\beta\gamma} \left( \1- \otheta \right) \label{connexsym}
\end{eqnarray}
where ${\Gamma^{\pm}}^{\alpha}_{\beta\gamma}$ denote the corresponding symbols on $V^\pm$ respectively.
These scalar distributions are associated to locally integrable functions given by
\begin{eqnarray}
{\Gamma}^{\alpha}_{\beta\gamma}= {\Gamma^{+}}^{\alpha}_{\beta\gamma} \theta +
{\Gamma^{-}}^{\alpha}_{\beta\gamma} \left( 1- \theta \right) . \label{connexsymfunc}
\end{eqnarray}
These functions may be discontinuous across $\S$ and, as in the general case (\ref{TonS}), we have
\begin{equation}
{\Gamma}^{\alpha}_{\beta\gamma}|_\S= \frac{1}{2} \left( {\Gamma^{+}}^{\alpha}_{\beta\gamma} |_\S +{\Gamma^{-}}^{\alpha}_{\beta\gamma} |_\S\right).\label{connex_at_S}
\end{equation}

Using now the standard formula (in a local coordinate basis) and treating the objects appearing as distributions we define the Riemann tensor distribution by
$$
\underline{R}^{\alpha}_{\beta\lambda\mu}=\partial_{\lambda}\underline{\Gamma}^{\alpha}_{\beta\mu}-
\partial_{\mu}\underline{\Gamma}^{\alpha}_{\beta\lambda}+\underline{\Gamma}^{\alpha}_{\lambda\rho}
\underline{\Gamma}^{\rho}_{\beta\mu}-\underline{\Gamma}^{\alpha}_{\mu\rho}\underline{\Gamma}^{\rho}_{\beta\lambda} .
$$
First of all, we observe that the products of $\underline\Gamma$'s are well defined because ${\underline{\Gamma}}^{\alpha}_{\beta\gamma}$ are distributions associated to functions, and actually they become (upon using $\otheta\cdot\otheta=\otheta$)
\begin{eqnarray*}
{\underline{\Gamma}}^{\alpha}_{\lambda\rho}{\underline{\Gamma}}^{\rho}_{\beta
\mu}={\Gamma^{+}}^{\alpha}_{\lambda\rho}{\Gamma^{+}}^{\rho}_{\beta\mu} \otheta + 
{\Gamma^{-}}^{\alpha}_{\lambda\rho}{\Gamma^{-}}^{\rho}_{\beta\mu} \left (\1-\otheta\right)
\end{eqnarray*}
On the other hand, we have from (\ref{connexsym}),  as in (\ref{nablaT1}):
\begin{eqnarray*}
\partial_{\mu}{\underline{\Gamma}}^{\alpha}_{\beta\lambda}=\partial_{\mu}{\Gamma^{+}}^{\alpha}_{\beta\lambda} \otheta + \partial_{\mu}{\Gamma^{-}}^{\alpha}_{\beta\lambda}  \left(\1-\otheta\right)
+ \left[ \Gamma^{\alpha}_{\beta\gamma} \right ]  n_{\mu} \deltaS
\end{eqnarray*}
so that the final expression for the Riemann tensor distribution reads
\be
{\underline{R}}^{\alpha}_{\beta\lambda\mu}= {R^{+}}^{\alpha}_{\beta\lambda\mu} \otheta + {R^{-}}^{\alpha}_{\beta\lambda\mu} \left(\1-\otheta\right)+\left(n_{\lambda}\left
[\Gamma^{\alpha}_{\beta\mu} \right ] - n_{\mu} \left [ 
\Gamma^{\alpha}_{\beta\lambda} \right ]\right)\deltaS \label{dRiem}
\ee
where ${R^{\pm}}^{\alpha}_{\beta\lambda\mu}$ are the Riemann tensors fields on $V^\pm$ respectively. 
The part proportional to $\deltaS$ is called the {\em  singular part of the Riemann tensor distribution}, and it has support only on $\S$. An explicit formula for this part in terms of the jump of the second fundamental form of $\S$ is provided in the main text, expression (\ref{HRie}). 

\section{Useful formulas}
\subsection{Concerning $\S$ and its objects}\label{App:D1}
Consider a hypersurface $(\S,h_{ab})$ imbedded in a n-dimensional spacetime $(V,g_{\a\b})$.
We will later use
this construction for the $+$ and $-$ sides.
Using the dual bases $\{n^\mu,e^\mu_a\}$ and $\{n_\mu,\omega^a_\mu\}$ introduced in Section \ref{sec:matching}, we have
\begin{eqnarray}
e^\r_a\N_\r e_b^\a &=& -K_{ab}n^\a + \overline{\Gamma}^c_{ab} e_c^\a,\\
e^\r_a\N_\r \omega^b_\a &=& -K_a^b n_\a - \overline{\Gamma}_{ac}^b \omega^c_\a,\\
e^\r_a\N_\r n_\a &=& K_{ab}\omega^b_\a
\end{eqnarray}
where $K_{ab}$ is the second fundamental form introduced in (\ref{2FF}) and 
$$
\overline{\Gamma}^c_{ab} \defi \omega^c_{\a}e^\r_a\N_\r e_b^\a
$$
represent the Christoffel symbols of the Levi-Civita connection associated to the first fundamental form $h_{ab}$ of $\S$. In general ---unless the jump of the second fundamental form (\ref{Kdisc}) vanishes--- there will be two versions, one $+$ and one $-$ of all these equations except for the last one, the connection, which is uniquely defined given that the first fundamental form agrees on both sides (\ref{h=h}). 

The covariant derivative defined by $\overline \Gamma$ is denoted by $\overline\nabla$. The relationship between $\nabla$ and $\overline\nabla$ on $\S$ is ruled by the following formula (given here for a $(1,1)$-tensor field $S^\b_\a$, but generalizable in the obvious way to arbitrary ranks \cite{MS})
\be
\omega^a_\b e^\a_be^\r_c \N_\r S^\b_\a = \overline{\N}_c \overline{S}^a_b + (e^\b_bS^\r_\b n_\r) K_{c}^a + (\omega^a_\a S^\a_\r n^\r) K_{cb} \label{nabla=nabla}
\ee
where, for any tensor field $S$, we denote by $\overline{S}$ its projection to $\S$:
\be
\overline{S}^a_b\defi \omega^a_\a e^\b_b S^\a_\b. \label{bar}
\ee
The equivalent ``space-time'' version of (\ref{nabla=nabla}) is
\be
h^\g_\b h^\a_\d h^\r_\sigma \N_\r S^\b_\a = \overline{\N}_\sigma \overline{S}^\g_\d + (h^\b_\d S^\r_\b n_\r) K_{\sigma}^\g + (h^\g_\a S^\a_\r n^\r) K_{\sigma\d} , \label{nabla=nabla1}
\ee
where $ \overline{S}^\g_\d$ is the spacetime version of $ \overline{S}^a_b$, i.e.
$ \overline{S}^\g_\d:=\omega^b_\delta e_a^\gamma\overline{S}^a_b=h^\g_\a h^\b_\d S^\a_\b$.

Denoting by $\Rb^d_{abc}$ the Riemann tensor of $(\S,h_{ab})$, the classical Gauss equation reads
\be
\omega^d_\a R^\a_{\b \g \d}e_a^\b e_b^\g e_c^\d = \Rb^d_{abc} - K_{ac}K_b^d + K_{ab}K^d_c, \label{gauss0}
\ee
whose contractions are
\begin{eqnarray}
e^\a_a e^\g_cR_{\a\g} - n^\a n^\g R_{\a \b\g\d}e_a^\b e_c^\d &=&\Rb_{ac} - K^d_d K_{ac} + K_{ab}K^b_c, \label{gauss2}\\
R - 2n^\a n^\b R_{\a\b} &=& \Rb - (K^d_d)^2 + K_{ab}K^{ab} \label{gauss3}
\end{eqnarray}
where $\Rb_{ac}$ and $\Rb$ denote the Ricci tensor and scalar curvature of $(\S,h_{ab})$.

Similarly, the classical Codazzi equation reads
\begin{equation}
n_\m R^{\m}_{\a\b\g}e_a^\a e^\b_b e^\g_c = \overline{\N}_c K_{ba} - \overline{\N}_b K_{ca} \label{codazzi}
\end{equation}
with contraction 
\be
n^\a R_{\a \g} e^\g_b = \overline{\N}_a K^a_b - \overline{\N}_b K^d_d .\label{codazzi2}
\ee

As mentioned before, generically there will be two versions of each of the previous equations, one for each $\pm$ side of $\S$ if this is a matching hypersurface. Thus, for instance (and using space-time notation), (\ref{gauss3}) and (\ref{codazzi2}) must have the two versions:
\bea
R^{\pm}-2R^{\pm}_{\mu\nu}n^\mu n^\nu = \Rb -(K^{\pm\rho}{}_\rho)^2+K^{\pm}_{\mu\nu}K^{\pm\mu\nu},\label{gauss}\\
n^{\mu}R^{\pm}_{\mu\rho}h^{\rho}{}_{\nu}=\overline\nabla^{\mu}K^{\pm}_{\mu\nu}-\overline\nabla_{\nu}K^{\pm\rho}{}_{\rho} . \label{coda}
\eea
On the other hand, equation (\ref{nabla=nabla}) at points of the \emph{matching} hypersurface $(\S,h_{ab})$
of the already glued spacetime $V=V^+\cup V^-$ reads 
\begin{equation}
h^\g_\b h^\a_\d h^\r_\sigma \N_\r S^\b_\a |_{\S}= \overline{\N}_\sigma \overline{S}^\g_\d + (h^\b_\d S^\r_\b n_\r) K^\S{}_{\sigma}^\g + (h^\g_\a S^\a_\r n^\r) K^\S_{\sigma\d},\label{nabla=nabla2}
\end{equation}
where  here we use $h\N|_\S$ just to make explicit that $h\N$ is
being restricted to points at $\S$ using the connection
given by (\ref{connexsymfunc}), whose restriction to $\S$ is (\ref{connex_at_S}).
Note that whenever both second fundamental forms coincide $[K_{\a\b}]=0$ on a
matching hypersurface $\S$, equation (\ref{nabla=nabla2}) reads just as (\ref{nabla=nabla1}).

\subsection{Discontinuities}\label{App:D2}
In the computations we need the discontinuities of objects, such as functions and tensor fields, across $\S$. This also implies that we need to control such discontinuities for the {\em derivatives} of those objects, and for their products. Here we provide the general rules.

Let $A$ and $B$ be any two functions possibly discontinuous across $\S$. Then
$$
\left[AB\right] = A^+ B^+ |_\S- A^- B^- |_\S = A^+ B^+ |_\S -A^+ B^- |_\S+ A^+ B^- |_\S- A^- B^- |_\S =A^+|_\S [B]+[A]B^-|_\S
$$
and an equivalent expression interchanging $A\leftrightarrow B$. Adding these two expressions and using (\ref{TonS}) we get
\be
\left[AB\right] =A^\S [B] + [A] B^\S . \label{discprod}
\ee 

Concerning derivatives, let us start with any function $f$ that may be discontinuous across $\S$. If we compute the tangent derivatives on both sides of $\S$ we obtain
$$
e^\mu_a\left[ \partial_\mu f\right]= \left[e^\mu_a \partial_\mu f\right]=e^\mu_a \partial_\mu f^+|_\S - e^\mu_a \partial_\mu f^-|_\S=\partial_a f^+|_\S -\partial_a f^+|_\S= \partial_a [f]=e^\mu_a\partial_\mu [f]
$$
and thus, by orthogonal decomposition,
\be
\left[\partial_\nu f\right] = F n_\nu + \omega^a_\nu e^\mu_a \partial_\mu [f] = F n_\nu + h^\mu_\nu \partial_\mu [f] \label{discf}
\ee
where $F$ is a function defined only on $\S$ that measures the discontinuity of the normal derivatives of $f$ across $\S$:
$$
F \defi n^\nu \left[\partial_\nu f\right] .
$$

Consider now the case of a one-form $t_\mu$, again possibly discontinuous across $\S$. A direct computation using (\ref{discf}) and (\ref{discprod}) produces
$$
e^\mu_a \left[\nabla_\mu t_\a\right] =e^\mu_a \left[\partial_\mu t_\a -t_\r\Gamma^\r_{\m\a}\right]=e^\mu_a\left(\partial_\mu\left[t_\a \right]-[t_\r] \Gamma^{\S \r}_{\m\a} \right)- t^\S_\r \left[\Gamma^\r_{\m\a} \right]e^\mu_a =e^\mu_a\nabla{}_\mu\left[t_\a \right]- t^\S_\r \left[\Gamma^\r_{\m\a} \right]e^\mu_a,
$$
Let us remark that the derivative $\nabla_{\vec{e}}$ is restricted to points on $\Sigma$, so that the connection (\ref{connex_at_S}) must be used. Therefore
\be
 \left[\nabla_\mu t_\a\right] =n_\mu T_\a +h^\n_\m \nabla{}_\n [t_\a]-t^\S_\r \left[\Gamma^\r_{\n\a} \right]h^\n_\m \label{disc1f_pre}
\ee
where $T_\a$ is a one-form defined only on $\S$ giving the discontinuity of the normal derivatives of $t_\a$ across $\S$,
$$
T_\a \defi n^\m  \left[\nabla_\mu t_\a\right],
$$
and the tangential derivative $h\nabla$ is restricted to $\S$, although it is not explicitly
indicated not to overwhelm the expressions.
The righthand side of (\ref{disc1f_pre}) can be further computed. First, due to (\ref{nabla=nabla2}) 
\bean
h^\n_\m \nabla{}_\n [t_\a] &=& \Nb_\mu \overline{[t_\a]}+n^\r [t_\r] K^\S_{\m\a}+n_\a n^\r h^\n_\m \nabla_\n [t_\r] \\
&=& \Nb_\mu \overline{[t_\a]}+n^\r [t_\r] K^\S_{\m\a}+n_\a \Nb_\m \left( [t_\r]n^\r\right) -n_\a [t^\r] K^\S_{\r\m}
\eean
while, for the last summand in (\ref{disc1f_pre}) we use (\ref{Gammadisc}) and (\ref{discK}) 
$$
-t^\S_\r \left[\Gamma^\r_{\n\a} \right]h^\n_\m =t^\S_\r n^\r [K_{\mu\a}]-n_\a t^\S_\r [K^\r_\m] .
$$
Introducing both results into (\ref{disc1f_pre}) we get
\begin{eqnarray}
&&\left[\nabla_\mu t_\a\right] =n_\mu T_\a +\Nb_\mu \overline{[t_\a]}+n^\r [t_\r] K^\S_{\m\a}+n_\a \left( \Nb_\m \left( [t_\r]n^\r\right) -[t^\r] K^\S_{\r\m}\right)+t^\S_\r n^\r [K_{\mu\a}]-n_\a t^\S_\r [K^\r_\m]\nonumber\\
&&\qquad=n_\mu T_\a +\Nb_\mu \overline{[t_\a]}+n^\r [t_\r K_{\m\a}]+n_\a \left( \Nb_\m \left( [t_\r]n^\r\right)-[t_\r K^\r_{\m}]\right).
\label{disc1f1_pre}
\end{eqnarray}
Observe that when there is no jump of the second fundamental form, $[K_{\a\b}]=0\, (\Leftrightarrow [\Gamma^\r_{\n\a}]=0)$, equations (\ref{disc1f_pre}) and (\ref{disc1f1_pre}) read, respectively,
\begin{eqnarray}
  &&\left[\nabla_\mu t_\a\right] =n_\mu T_\a +h^\n_\m \nabla_\n [t_\a],\label{disc1f}\\
  &&\left[\nabla_\mu t_\a\right] =n_\mu T_\a +\Nb_\mu \overline{[t_\a]}+n^\r [t_\r] K_{\m\a}+n_\a \left( \Nb_\m \left( [t_\r]n^\r\right) -[t^\r] K_{\r\m}\right)\label{disc1f1}.
\end{eqnarray}
These formulas can be generalized to arbitrary $(p,q)$-tensor fields $T^p _q$ in an obvious way. In that case, the term replacing $T_\a$ is simply a tensor field of the same type (and with the same symmetry and trace properties) as $T^p_q$, defined only on $\S$ and measuring the discontinuities of the normal derivatives of $T^p_q$.

\subsection{Derivatives of tensor distributions proportional to $\deltaS$}\label{App:D3}
Let us consider tensor distributions of type
$$
t_{\a_1\dots \a_p} \deltaS
$$
where $t_{\a_1\dots \a_p}$ is any tensor field defined {\em at least} on $\S$, but not necessarily off $\S$ (for instance $h_{\m\n}$ or $n_\m$ are not defined outside $\S$). We want to compute the covariant derivative of such tensor distributions. Then we have
\bean
\left< \nabla_\l \left(t_{\a_1\dots \a_p}  \deltaS\right),Y^{\l\a_1\dots \a_p}\right>=
-\left<t_{\a_1\dots \a_p} \deltaS, \nabla_\l Y^{\l\a_1\dots \a_p} \right>=
-\left<\deltaS, t_{\a_1\dots \a_p}  \nabla_\l Y^{\l\a_1\dots \a_p} \right>\\
=-\int_\S t_{\a_1\dots \a_p}  \nabla_\l Y^{\l\a_1\dots\a_p} d\sigma =-\int_\S t_{\a_1\dots \a_p}  (n_\l n^\r +h^\r_\l) \nabla_\r Y^{\l\a_1\dots\a_p} d\sigma .
\eean
The first summand here is
$$
-\left<t_{\a_1\dots \a_p}  n_\l n^\r \deltaS, \nabla_\r Y^{\l\a_1\dots\a_p} \right>=
\left<\nabla_\r\left(t_{\a_1\dots \a_p}  n_\l n^\r \deltaS\right),  Y^{\l\a_1\dots\a_p} \right>
$$
while the second one has derivatives tangent to $\S$ and thus 
\bean
-\int_\S t_{\a_1\dots \a_p}   h^\r_\l \nabla_\r Y^{\l\a_1\dots\a_p} d\sigma = 
-\int_\S   h^\r_\l \nabla_\r (t_{\a_1\dots \a_p}  Y^{\l\a_1\dots\a_p}) d\sigma +
\int_\S  Y^{\l\a_1\dots\a_p} h^\r_\l \nabla_\r t_{\a_1\dots \a_p}  d\sigma 
\eean
and using (\ref{nabla=nabla1}) for the first integral here
\bean
&=&  -\int_\S   \Nb_\l (\overline{t_{\a_1\dots \a_p}  Y^{\l\a_1\dots\a_p}}) d\sigma -
\int_\S K^{\S\r}{}_\r \, n_\l t_{\a_1\dots \a_p}  Y^{\l\a_1\dots\a_p} d\sigma +
\int_\S  Y^{\l\a_1\dots\a_p} h^\r_\l \nabla_\r t_{\a_1\dots \a_p}  d\sigma\\
&=& \left< \left( h^\r_\l \nabla_\r t_{\a_1\dots \a_p} -K^{\S\r}{}_\r \, n_\l t_{\a_1\dots \a_p}\right)\deltaS, Y^{\l\a_1\dots\a_p} \right> \hspace{2cm} 
\eean
where we have used that, as $Y^{\l\a_1\dots\a_p}$ has compact support, the first total divergence term integrates to zero.
Summing up, we have the following basic formula
\be
\nabla_\l \left(t_{\a_1\dots \a_p}  \deltaS \right) =\nabla_\r\left(t_{\a_1\dots \a_p}  n_\l n^\r \deltaS\right)+
 \left( h^\r_\l \nabla_\r t_{\a_1\dots \a_p} -K^{\S\r}{}_\r \, n_\l t_{\a_1\dots \a_p}\right)\deltaS \label{nablaTdelta} .
\ee
In particular, for the second derivative of $\otheta$ one gets
\be
\nabla_\n\nabla_\m \otheta=\nabla_\n (n_\m \deltaS) = \nabla_\rho (n_\m n_\n n^\rho \deltaS) + \left(K_{\m\n} -K^{\S\r}{}_\r n_\m n_\n \right)\deltaS \label{nablandelta} .
\ee

\subsection{Ricci and Bianchi identities}\label{App:D4}
The Bianchi identity holds in the distributional sense, for a proof consult \cite{MS}:
\begin{equation}
\label{eq:Bianchi_id}
\N_\r \underline R_{\a\b\n\m} + \N_\n \underline R_{\a\b\m\r} + \N_\m \underline R_{\a\b\r\n} = 0.
\end{equation}

Concerning the Ricci identity, let us consider a one-form which may have a discontinuity across $\S$. It can be written as 1-covariant tensor and as a one-form distribution as
$$
t_\a =t^+_\a \theta + t^-_\a (1-\theta) ; \hspace{1cm} \underline{t}_\a =t^+_\a \otheta + t^-_\a (\1-\otheta) 
$$
To compute the derivatives, we need to take $\underline{t}_\a$ as a distribution. Then, from (\ref{nablaT1}) we first have
$$
\nabla_\m \underline{t}_\a =\nabla_\m t^+_\a \otheta +\nabla_\m t^-_\a (\1 -\otheta) +[t_\a] n_\mu \deltaS
$$
and applying (\ref{nablaT1}) to the first part not proportional to $\deltaS$ we derive
\be
\nabla_\l \nabla_\m \underline{t}_\a = \nabla_\l \nabla_\m t^+_\a \otheta +\nabla_\l \nabla_\m t^-_\a (\1 -\otheta) +
[\nabla_\m t_\a] n_\l \deltaS +\nabla_\l \left([t_\a] n_\mu \deltaS\right) . \label{nabla2t}
\ee
Formula (\ref{nablaTdelta}) gives the last term here
$$
\nabla_\l \left([t_\a] n_\mu \deltaS\right)=\N_\r \left([t_\a] n_\mu n_\l n^\r \deltaS \right)+
 \left(n_\mu h^\r_\l \nabla_\r [t_\a] +  [t_\a] K_{\l\m} -K^{\S\r}{}_\r \, n_\l[t_\a] n_\mu\right)\deltaS .
$$
Introducing (\ref{disc1f_pre}) into (\ref{nabla2t}) and using this last result we arrive at
\bean
(\nabla_\l \nabla_\m -\nabla_\m \nabla_\l) \underline{t}_\a &=& (\nabla_\l \nabla_\m -\nabla_\m \nabla_\l) t^+_\a \, \otheta +
(\nabla_\l \nabla_\m -\nabla_\m \nabla_\l)t^-_\a (\1 -\otheta) \\
&& -t^\S_\r  \left(n_\l  [\Gamma^\r_{\m\a}]-n_\m [\Gamma^\r_{\l\a}] \right)\deltaS
\eean
and using here the Bianchi identity on both $\pm$ regions and expression (\ref{Riedist}) we finally get
\be
(\nabla_\l \nabla_\m -\nabla_\m \nabla_\l) \underline{t}_\a = -t_\r \underline{R}^\r_{\a\l\m} .
\ee
Of course, this can be extended to tensor fields of any $(p,q)$ type which may have discontinuities across $\S$. 

What about the Ricci identity for tensor distributions not associated to tensor fields? The answer now is much more involved, and it must be treated case by case, because taking covariant derivatives presents several problems. As an illustrative example, let us analyze the case of the second covariant derivative of $\deltaS$. For the first derivative we have from (\ref{nablaTdelta})
\be
\nabla_\mu \deltaS = \N_\r\left(n_\m n^\r \deltaS\right)-K^{\S\r}_\r n_\mu  \deltaS \label{nabladelta}
\ee
so that defining a one-form distribution $\Delta_\mu$ with support on $\S$ as follows
\be
\left<\Delta_\mu ,Y^\mu\right> \defi -\int_\S  n_\m  n^\r\nabla_\r Y^\m d\sigma ; \hspace{1cm} \Delta_\m = \N_\r\left(n_\m n^\r \deltaS\right)\label{deltaprime}
\ee
we can also write
$$
\nabla_\mu \deltaS = \Delta_\m -K^{\S\r}_\r n_\mu  \deltaS .
$$
Note, however, that $\Delta_\m$, and therefore $\nabla_\m \deltaS$ too, is only well defined when acting on test vector fields whose {\em covariant} derivative is locally integrable on $\S$. Thus, the second covariant derivative of $\deltaS$ is not defined in the general case with a discontinuous connection $\Gamma^\a_{\m\n}$. To see this, observe that to define $\nabla_\l \nabla_\m \deltaS$ we need to define $\nabla_\l \Delta_\m$, but this should be according to definition \ref{derivative} 
\be
\left< \nabla_\l \Delta_\m , Y^{\l\m}\right >=-\left<  \Delta_\m , \nabla_\l Y^{\l\m}\right >\label{nablaDelta}
\ee
and this is ill-defined because $\nabla_\l Y^{\l\m}$ does {\em not} have a locally integrable covariant derivative in the sense of functions: actually, its covariant derivative can only be defined in the sense of distributions.

Nevertheless, if the connection is continuous, that is, $[\Gamma^\a_{\m\n}]=0$, then (\ref{nablaDelta}) makes perfect sense because the covariant derivative $\nabla_\r\nabla_\l Y^{\l\m}$ is a locally integrable tensor field. Thus, in this case we can write
\be
\left< \nabla_\l \Delta_\m , Y^{\l\m}\right >=\int_\S n_\m n^\r \nabla_\r \nabla_\l Y^{\l\m} d\sigma 
\ee
and we can prove the Ricci identity for distributions such as $\deltaS$. To that end, a straightforward if somewhat lengthy calculation, using the Ricci identity under the integral and the rest of techniques hitherto explained, leads to the following explicit expression:
\bean
\N_\l\N_\m \deltaS&=& \N_\r \N_\sigma (n_\m n_\l n^\r n^\sigma \deltaS)+
\N_\r\{(K_{\l\m}-K^\sigma_\sigma n_\l n_\m) n^\r \deltaS\}+\\
&&\deltaS\{K^\sigma_\sigma(K^\r_\r n_\l n_\m -K_{\l\m})+n^\r n^\sigma R^\S_{\r\m\l\sigma}+K_\l^\r K_{\m\r}+n_\l n_\m K^{\r\sigma}K_{\r\sigma} \}+\\
&&\deltaS n_\m \{\Nb_\r K^\r_\l-\Nb_\l K^\r_\r-n^\r R^\S_{\r\l} \}
\eean
where all the summands are obviously symmetric in $(\l\m)$ except for those in the last line which, by virtue of the contracted Codazzi relation (\ref{coda}), become simply
$n^\r n^\sigma R^\S_{\r\sigma} n_\l n_\m \deltaS$, so that finally one arrives at the desired result
$$
\N_\l\N_\m \deltaS -\N_\m\N_\l \deltaS =0 .
$$

\section{Problems with the use of Gaussian coordinates based on the matching hypersurface $\S$}\label{App:E}
In the literature on junction conditions \cite{DSS} or in general when dealing with braneworlds, it is customary to simplify the difficulties of dealing with tensor distributions by using Gaussian coordinates based on the matching hypersurface and  a classical Dirac delta ``function''. This leads to some subtleties very often ignored and, in fact,  to unsolvable problems if one is to describe gravitational double layers.
In this Appendix we clarify this situation and provide a useful translation between the rigorous and the simplified methods.
Choose local Gaussian coordinates $\{y,u^a\}$ based on the matching hypersurface $\S$ given by 
$$
\S : \hspace{3mm} \{y=0\}
$$
so that the metric reads locally around $\S$ as
$$
ds^2 = dy^2 +g_{ab}(y,u^c)dx^adx^b .
$$
We can identify the local coordinates of $\S$ as $\xi^a=u^a$, or in other words, the parametric representation of $\S$ and the tangent vector fields $\vec{e}_a$ are simply
$$
\{y=0, u^a= \xi^a\},  \hspace{1cm} \vec{e}_a =\left.\frac{\partial}{\partial u^a}\right|_{y=0}. 
$$
The unit normal is in this case
$$
\bm{n} = dy |_{y=0} 
$$
and the first fundamental form (\ref{h=h}) becomes simply 
$$
h_{ab}=g_{ab}(0,u^c).
$$
In what follows, $h$ denotes the determinant of $h_{ab}$.
The two regions matched are represented by $y>0$ and by $y<0$. A trivial calculation proves that the second fundamental forms inherited from both sides are
$$
K^\pm_{ab} =\lim_{y\rightarrow 0^\pm} \partial_y g_{ab} \hspace{1cm} \Longrightarrow  \hspace{1cm} \left[K_{ab}\right] = \left[\partial_y g_{ab}\right]|_{y=0} .
$$

In these coordinates, the $\S$-step function (\ref{theta}) can be easily identified with the standard Heaviside step function $\theta(y)$. Thus, its covariant derivative is easily computed
$$
\nabla \theta(y) = \delta(y) dy
$$
where $\delta(y)$ is the Dirac delta ``function''. This can be immediately put in correspondence with (\ref{nablatheta}) in such a way that, in this coordinate system
$$
\deltaS \leftrightarrow \delta(y) \, .
$$
Now, if we multiply $\delta(y)$ for any function then 
$$
F\delta(y) = F|_{y=0} \delta(y) \leftrightarrow F\deltaS = F|_\S \deltaS .
$$
Observe, however, that a first subtlety arises: when we apply $\delta(y)$ to any test function $Y(x^\m)$, we do not simply get $Y|_{y=0}$, but we also need to integrate on $\S$, that is 
$$
\left<\delta(y), Y \right> =\int_{y=0} Y(y=0,u^c) \sqrt{-h}\,  du^{n}.
$$
This corresponds to (\ref{deltaS}). 

The discontinuity of the connection (\ref{Gammadisc}) together with (\ref{discK}) can be expressed by giving the non-zero jumps of the Christoffel symbols
$$
[\Gamma^y_{ab}] = -[K_{ab}], \hspace{1cm} [\Gamma^a_{by}] = [K^a{}_b] 
$$
and similarly (\ref{HRie}), (\ref{HRic}) 
and (\ref{HG}) read (only the non-zero components shown)
\bean
H_{yayb}=-[K_{ab}], \hspace{1cm} H_{yy} =-[K^c_c], \,\,\, H_{ab} = -[K_{ab}], 
\hspace{1cm} {\cal G}_{ab} =-[K_{ab}] +[K^c_c] h_{ab} 
\eean
so that, for instance, the Einstein tensor tangent components acquire a term proportional to $\delta(y)$ given by ${\cal G}_{ab} \delta(y)$. 

And now is when the real problems start: if one is to compute covariant derivatives of the curvature tensors, or the Einstein tensor, as distributions, one needs to compute terms such as, say,  $\nabla_\m ({\cal G}_{ab} \delta(y))$. How to do that? 
Even simpler, how to compute $\nabla_\mu \delta(y)$? One might naively write
$$
\nabla \delta(y)\,  \xcancel{=} \,  \delta'(y) dy \hspace{2cm} \mbox{wrong!}
$$
where $\delta'(y)$ is ``the derivative" of the Dirac delta. This is clearly ill-defined, because one does not know how such a $\delta'(y)$ should act on test functions (as minus the integral on $\S$ of the $y$-derivative of the test function?). But worse, even if one could find a proper definition of such a $\delta'(y)$, still the formula would miss the second essential term appearing in (\ref{nabladelta}) which is proportional to $\deltaS$ and {\em depends on the extrinsic properties of the matching hypersurface} via the trace of its second fundamental form.

In order to show how to proceed if one insists in using Gaussian coordinates, the computation of $\nabla\delta(y)$ must go as follows (here $g$ stands for the determinant of $g_{\a\b}$) 
\bean
\left<\nabla\delta(y), \vec Y \right>= -\left<\delta(y),\nabla_\m Y^\m\right> =-\int_{y=0}\nabla_\m Y^\m du^n= -\int_{y=0} \frac{1}{\sqrt{-g}}\partial_\m(\sqrt{-g}Y^\m)du^n\\
=-\int_{y=0} \frac{1}{\sqrt{-h}}\partial_\m(\sqrt{-h}Y^\m)du^n=-\int_{y=0}
\left(\partial_y Y^y +\partial_a Y^a+Y^\m  \frac{1}{\sqrt{-h}}\partial_\m\sqrt{-h}\right)du^n \\
=-\int_{y=0}
\left(\partial_y Y^y +\frac{1}{\sqrt{-h}}\partial_a (\sqrt{-h}Y^a) +Y^y \frac{1}{\sqrt{-h}}\partial_y\sqrt{-h}\right)du^n=-\int_{y=0}
\left(\partial_y Y^y +Y^y K^\S{}^a_a \right)du^n.
\eean
In the last step we have used Gauss' theorem. This formula corresponds to (\ref{nabladelta}). Observe the fact that the extrinsic curvature $K_{ab}$ is
not necessarily equal as computed from either side of $y=0$
and therefore it is not univocally defined. Hence, a definite prescription of what is
its value on $\S$, that is $K^\S_{ab}$, must be provided.

The above subtleties and difficulties when using Gaussian coordinates are probably the reasons why double layers were not found in quadratic $F(R)$ or other quadratic theories until they were derived in \cite{Senovilla13,Senovilla14,Senovilla15}.

\bibliographystyle{review_bib.bst}	
\bibliography{references.bib}

\end{document}